\documentclass[journal]{IEEEtran}
\usepackage{amsmath,amsfonts}
\usepackage{algorithmic}
\usepackage{algorithm}
\usepackage{array}
\usepackage{lipsum}
\ifCLASSOPTIONcompsoc
\fi
\usepackage{subfig}
\usepackage{caption}
\usepackage{textcomp}
\usepackage[export]{adjustbox}
\usepackage{stfloats}
\usepackage{url}
\usepackage{hhline}
\usepackage{verbatim}
\usepackage{makecell}
\usepackage{graphicx}
\usepackage{cite}
\usepackage{physics}
\usepackage{CJKutf8}
\usepackage{multirow}

\begin{document}
\bstctlcite{IEEEexample:BSTcontrol}
\title{Personalized Adversarial Data Augmentation for Dysarthric and Elderly Speech Recognition}

\author{Zengrui Jin, Mengzhe Geng, Jiajun Deng, Tianzi Wang, Shujie Hu, Guinan Li, Xunying Liu 
	\thanks{Zengrui Jin, Mengzhe Geng, Jiajun Deng, Tianzi Wang, Shujie Hu, Guinan Li, Xunying Liu are with the Chinese University of Hong Kong, China (email: \{zrjin,mzgeng,jjdeng,twang,sjhu,gnli,xyliu\}@se.cuhk.edu.hk).\\
	}
}




\maketitle

\begin{abstract}
Despite the rapid progress of automatic speech recognition (ASR) technologies targeting normal speech, accurate recognition of dysarthric and elderly speech remains highly challenging tasks to date. 
It is difficult to collect large quantities of such data for ASR system development due to the mobility issues often found among these users. 
To this end, data augmentation techniques play a vital role. In contrast to existing data augmentation techniques only modifying the speaking rate or overall shape of spectral contour, fine-grained spectro-temporal differences between dysarthric, elderly and normal speech are modelled using a novel set of speaker dependent (SD) generative adversarial networks (GAN) based data augmentation approaches in this paper.
These flexibly allow both: 
a) temporal or speed perturbed normal speech spectra to be modified and closer to those of an impaired speaker when parallel speech data is available; and 
b) for non-parallel data, the SVD decomposed normal speech spectral basis features to be transformed into those of a target elderly speaker before being re-composed with the temporal bases to produce the augmented data for state-of-the-art TDNN and Conformer ASR system training. 
Experiments are conducted on four tasks: the English UASpeech and TORGO dysarthric speech corpora; the English DementiaBank Pitt and Cantonese JCCOCC MoCA elderly speech datasets. 
The proposed GAN based data augmentation approaches consistently outperform the baseline speed perturbation method by up to 0.91\% and 3.0\% absolute (9.61\% and 6.4\% relative) WER reduction on the TORGO and DementiaBank data respectively. 
Consistent performance improvements are retained after applying LHUC based speaker adaptation.
\end{abstract}

\begin{IEEEkeywords}
Dysarthric Speech; Elderly Speech; Speech Recognition; Data Augmentation; GAN; hybrid TDNN; end-to-end Conformer
\end{IEEEkeywords}

\section{Introduction}
\label{sec:intro}

Despite the rapid progress of automatic speech recognition (ASR) technologies targeting normal speech in recent decades \cite{bahl1986maximum, graves2013speech, peddinti2015time, povey2016purely, chan2016listen, wang2020transformer, gulati2020conformer}, accurate recognition of pathological and elderly speech remains highly challenging tasks to date \cite{christensen2012comparative, christensen2013combining, sehgal2015model, yu2018development, hu2019cuhk, liu2020exploiting, ye2021development, geng2021investigation}. 
Dysarthria is a common form of speech disorder caused by a range of motor control conditions including cerebral palsy, amyotrophic lateral sclerosis, stroke and traumatic brain injuries \cite{whitehill2000speech, makkonen2018speech, scott1983speech, jerntorp1992stroke, lanier2010speech}. 
In a wider context, speech and language impairments are also commonly found among older adults experiencing natural ageing and neurocognitive disorders, for example, Alzheimer's disease \cite{fraser2016linguistic, alzheimer2021}. 
People with speech disorders often experience co-occurring physical disabilities and mobility limitations.
Their difficulty in using keyboard, mouse and touch screen based user interfaces makes voice based assistive technologies more natural alternatives \cite{hux2000accuracy, young2010difficulties} even though speech quality is degraded.
To this end, in recent years there has been increasing interest in developing ASR technologies that are suitable for dysarthric \cite{christensen2013combining, vachhani2017deep, kim2018dysarthric, joy2018improving, liu2019exploiting, shor2019personalizing, lin2020staged, kodrasi2020spectro, xiong2020source, liu2021recent, xie2022variational, takashima2020two, hermann2020dysarthric, wang2021improved, wang2021study, macdonald2021disordered, green2021automatic} and elderly speech \cite{vipperla2010ageing, rudzicz2014speech, zhou2016speech, konig2018fully, toth2018speech, ye2021development, pan2021using}.

Dysarthric and elderly speech bring challenges on all fronts to current deep learning based automatic speech recognition technologies predominantly targeting normal speech recorded from healthy, non-aged users.
First, a large mismatch between such data and healthy, normal voice is often observed.
Such difference manifests itself on many fronts including articulatory imprecision, decreased volume and clarity, changes in pitch, increased dysfluencies and slower speaking rate.
Second, the co-occurring disabilities and mobility limitations often found among impaired and elderly speakers lead to the difficulty in collecting large quantities of such data that are essential for current data intensive deep learning based ASR system development. 
Prior researches on dysarthric and elderly speech corpus development are mainly conducted for English \cite{menendez1996nemours, kim2008dysarthric, choi2011design, rudzicz2012torgo, macdonald2021disordered}.
A set of publicly available disordered and elderly speech corpora are shown in Tab. \ref{tab:datasets}.

Among these, the Nemours \cite{menendez1996nemours} corpus contains less than 3 hours of speech from 11 speakers, while the 15-hour TORGO \cite{rudzicz2012torgo} dataset is moderately larger. 
By far, the largest available and widely used dysarthric speech database, the English UASpeech \cite{kim2008dysarthric} corpus, contains 102.7 hours of speech recorded from 29 speakers among whom 16 are dysarthric speakers while the remaining 13 are healthy speakers. 
As the largest publicly available elderly speech dataset designed for neurocognitive disorder screening, the DementiaBank Pitt \cite{becker1994natural} corpus contains approximately 33 hours of audio data recorded over interviews between the 292 elderly participants and the clinical investigators.
Compared with more widely available normal speech corpora, such as Switchboard and Fisher conversational telephone speech \cite{godfrey1992switchboard} or LibriSpeech \cite{panayotov2015librispeech} containing from hundreds to thousands of hours of audio data, existing dysarthric and elderly speech corpora are much smaller in size. 
Similar data scarcity is also found not only among dysarthric speech datasets collected for non-English languages such as Dutch \cite{yilmaz2016dutch}, Italian \cite{turrisi2021easycall}, Cantonese \cite{wong2015development} and Korean \cite{choi2011design}, but also in elderly speech corpora. 

In addition, sources of variability commonly found in normal speech including accent or gender, when further compounded with those over age as well as speech and language pathology severity, create large diversity among dysarthric and elderly speakers \cite{mengistu2011adapting,xie2019fast}. 
Their mobility issues further limit the amount of speaker-level data available for fine-grained model based adaptation to facilitate user personalization of ASR systems \cite{xiong2020source,geng2021investigation,ye2021development,liu2021recent,takashima2020two,shor2019personalizing,green2021automatic}.

\begin{table}[t]
	\centering
	\caption{Description of publicly available dysarthric and elderly speech corpora for English (ENG.), Cantonese (CAN.), Italian (IT.) and Dutch (NL.).}
	\label{tab:datasets}
    \renewcommand\tabcolsep{2.5pt}
	\scalebox{1}{
	\begin{tabular}{l|c|c|c|c|c}
	  \hline
	  \hline
	  Corpus & Type & Lang. & \# Hours & \# Spkr. & \# Vocab. \\
	  \hline
	  \hline
	  Nemours \cite{menendez1996nemours} & \multirow{3}{*}{Dysarthric} & \multirow{3}{*}{ENG.} & 3 & 11 & -  \\
	  UASpeech \cite{kim2008dysarthric} &  &  & 102.7 & 29 & 455  \\
	  TORGO \cite{rudzicz2012torgo} &  &  & 15 & 15 & 1573  \\
	  \hline
	  EasyCall \cite{turrisi2021easycall} & \multirow{3}{*}{Dysarthric} & IT. & - & 55 & \multirow{3}{*}{-} \\
	  EST \cite{yilmaz2016dutch} &  & NL. & 6.3 & 16 &   \\
	  CUDYS \cite{wong2015development} &  & CAN. & 10 & 16 & \\
	  \hline
	  \hline
	  DementiaBank Pitt \cite{becker1994natural} & \multirow{2}{*}{Elderly} & ENG. & 33.1 & 688 & 3.8k  \\
	  JCCOCC MoCA \cite{xu2021speaker} &  & CAN. & 52 & 369 & 610k \\
	  \hline
	  \hline
	\end{tabular}
	}
\end{table}

To this end, data augmentation techniques play a vital role to address the above data sparsity issue. 
Data augmentation techniques have been widely studied in terms of normal speech recognition tasks. 
To expand the limited training data, tempo, speed or vocal tract length perturbation (VTLP) \cite{verhelst1993overlap, kanda2013elastic, jaitly2013vocal, ko2015audio}, stochastic feature mapping \cite{cui2015data}, SpecAugment \cite{park19e_interspeech}, cross domain feature adaptation \cite{bell2012transcription}, simulating noisy and reverberant speech \cite{ko2017study}, and back translation based methods for end-to-end systems \cite{hayashi2018back}, have been proposed and the improvement on coverage of the expanded training data leads to the improvement of generalization for ASR systems. 

In contrast, only limited research on data augmentation methods targeting dysarthric and elderly speech has been conducted. 
Inspired by the spectro-temporal level differences between impaired speech and normal speech such as overall reduction of speech volume, changes in the spectral envelope shape, weakened formants and slower speaking rate, recent research in this direction has been largely focused on front-end signal processing based techniques including tempo-stretching \cite{vachhani2018data, xiong2019phonetic}, VTLP \cite{jaitly2013vocal}, and speed perturbation \cite{ko2015audio, geng2021investigation}, of normal speech recorded from healthy control speakers.
The resulting speech data exhibiting certain high-level attributes such as a slower speaking rate and reduced speech volume is then used to augment the limited dysarthric or elderly speech training data \cite{geng2021investigation, liu2021recent, ye2021development}. 
A range of speech augmentation approaches investigated for dysarthric speech recognition \cite{geng2021investigation} suggest the combined use of personalized, speaker dependent (SD) together with speaker independent (SI) speed perturbation factors produces the largest performance improvements. 

One issue associated with the existing data augmentation approaches, for example, tempo or speed perturbation, is that fine-grained differences between normal and impaired speech are not fully considered during the augmentation process. 
Although the overall decrease in speaking rate and speech volume can be characterized using speed perturbation, other more detailed and prominent features associated with dysarthric and elderly speech including articulatory imprecision, decreased vocal clarity, breathy and hoarse voice cannot be fully accounted for. 
In order to address this issue, one general solution considered in this paper is to model the fine-grained spectro-temporal differences between dysarthric, elderly and normal speech using generative adversarial networks (GANs) \cite{goodfellow2014generative, mirza2014conditional, radford2015unsupervised} during data augmentation. 

In recent years GANs have been successfully applied to a wide range of normal speech processing tasks including, but not limited to: 
1) speech synthesis \cite{you2021gan, beck2022wavebender};
2) voice conversion \cite{lee2020many, wang2021enriching};
3) speech enhancement \cite{liu2020cp, su2021hifi};
4) code-switching sentence generation \cite{chang2019code, li2021improving};
5) speech emotion recognition \cite{su2021conditional, ma2022data};
6) speaker verification \cite{zhang2018vector, liu2020text, kataria2021deep} and
7) speech recognition \cite{zhao2019multi, du2020double, haidar2021fine}.

Prior researches on GAN based data augmentation for ASR applications targeting normal speech include synthesizing noisy speech to enhance the environmental robustness \cite{sheng2018data, hu2018generative, qian2019data, chen2022noise};
generating data for whisper recognition \cite{gudepu2020whisper}, speech translation \cite{mccarthy2020skinaugment}, far-field recognition \cite{mirheidari2020data} and voice search \cite{chen2020improving}. 
GAN based data augmentation approaches have also been studied in the context of speech emotion recognition \cite{chatziagapi2019data,tiwari2020multi,eskimez2020gan, bao2019cyclegan,yi2019adversarial,ma2022data} and speaker verification \cite{yang2018generative, chien2019bayesian, wang2020data, du2021synaug, nidadavolu2019cycle, shahnawazuddin2020domain}.

In contrast, only a few prior researches on applying GAN to dysarthric speech processing tasks, represented by dysarthric speech augmentation \cite{harvill2021synthesis, jin21_interspeech, huang2021towards, soleymanpour2022synthesizing} and dysarthric speech reconstruction \cite{chen2018generative, wang2022speaker} have been conducted to date. 
To the best of our knowledge, no previous study on GAN based data augmentation approaches designed for elderly speech recognition tasks has been conducted. 
One prominent factor that hinders the application of GAN to elderly speech lies in the precise nature of the underlying task specific data collection protocol. 
The majority of the disordered speech datasets \cite{macdonald2021disordered, menendez1996nemours, kim2008dysarthric, rudzicz2012torgo, wong2015development, choi2011design, yilmaz2016dutch, turrisi2021easycall} are designed using parallel speech recordings of identical spoken contents but different duration to facilitate pathological severity assessment by professional speech therapists.
In contrast, the elderly speech datasets represented by the English DementiaBank Pitt corpus \cite{becker1994natural} are constructed using a different protocol that is based on non-parallel, spontaneous conversational speech recorded during neurocognitive assessment interviews between elderly participants and clinical investigators.
In order to address the above issues, a novel set of generative adversarial networks (GAN) based speaker dependent, personalized data augmentation approaches tendering for the detailed characteristics of dysarthric and elderly speech recognition tasks discussed above are proposed in this paper. 

First, for dysarthric speech data where the underlying spoken contents are constrained to be parallel but the speech duration is allowed to vary, speaker dependent deep convolutional GANs (DCGANs) are trained to transform tempo or speed perturbed healthy speech spectra into those of individual target dysarthric speakers. 
The resulting data augmentation process not only allows the overall change in speaking rate, speech volume and spectral envelope to be simulated as in classic front-end level tempo or speed perturbation, but also ensures further fine-grained spectro-temporal characteristics associated with impaired speech including articulatory imprecision, decreased clarity, breathy and hoarse voice as well as increased dysfluencies and pauses to be injected into the augmented data. 

Second, to account for the non-parallel nature of elderly or dysarthric speech datasets, SVD based speech spectrum decomposition is used to derive structured speech representations: 
a) time invariant spectral basis features more closely related to speaker characteristics, for example, an average description of speech volume and energy distribution over frequency components; and 
b) time variant temporal basis features more related to spoken contents. 
The resulting spectral basis matrix of a healthy, non-aged source speaker is then transformed using speaker dependent spectral basis perturbation GANs into those of a target elderly or dysarthric speaker, before being re-composed with the source speaker's temporal basis to produce the final personalized augmented data for state-of-the-art hybrid TDNN \cite{peddinti2015time} and end-to-end (E2E) Conformer \cite{gulati2020conformer} ASR system training. 

Experiments are conducted on four tasks across two languages: the English UASpeech \cite{kim2008dysarthric} and TORGO \cite{rudzicz2012torgo} dysarthric speech corpora, the English DementiaBank Pitt \cite{becker1994natural} and Cantonese JCCOCC MoCA \cite{xu2021speaker} elderly speech datasets. 
Among these, the UASpeech and DementiaBank Pitt corpora are the largest publicly available datasets on dysarthric speech and elderly speech respectively. 
The proposed parallel and non-parallel GAN based dysarthric and elderly data augmentation approaches consistently outperform the baseline speed perturbation \cite{ko2015audio} and SpecAugment \cite{park19e_interspeech} methods by up to 0.91\% and 3.0\% absolute (9.61\% and 6.4\% relative) reduction in word error rate (WER). 
Consistent performance improvements are retained after model based speaker adaptation using learning hidden unit contribution (LHUC) is further applied.

The main contributions of this paper are summarized below:

1. To the best of our knowledge, this paper presents the first work to systematically investigate adversarial learning based data augmentation for dysarthric and elderly speech recognition tasks.
In contrast, existing data augmentation studies for dysarthric and elderly speech mainly focus on using signal-level tempo or speed perturbation based methods \cite{geng2021investigation, liu2021recent,vachhani2018data, xiong2019phonetic, ye2021development}. 
The only previous research on GAN based dysarthric speech augmentation required the explicit use of parallel speech data of the UASpeech corpus \cite{harvill2021synthesis, jin21_interspeech}. 
The non-parallel GAN based normal to pathological voice conversion approach studied in \cite{huang2021towards} is evaluated on naturalness and severity, but not measured in terms of the performance of ASR systems constructed using the generated data. 
No GAN based data augmentation approaches designed for spontaneous and conversational elderly speech recognition tasks using non-parallel data have been published to date.

2. The proposed spectral basis GAN model benefits from a distinct advantage of disentangling the time invariant speaker specific spectral characteristics from time variant temporal features that are more related to spoken contents. 
This novel approach broadens the application scope of GAN based data augmentation, and allows them to be flexibly performed on both parallel and non-parallel dysarthric or elderly speech data for ASR system development. 

3. The proposed adversarial data augmentation approaches achieved statistically significant performance improvements over the baseline hybrid TDNN and E2E Conformer systems using state-of-the-art data augmentation techniques including speaker independent and speaker dependent speed perturbation as well as SpecAugment, by up to  0.91\% and 3.0\% absolute (9.61\% and 6.4\% relative) word error rate (WER) reduction on four dysarthric or elderly speech recognition tasks across two languages. 
These findings serve to demonstrate the efficacy and genericity of our proposed GAN based data augmentation methods for dysarthric and elderly speech recognition. 

The rest of the paper is organized as follows.
Traditional data augmentation approaches are presented in Sec. \ref{sec:traditional_da}.
Adversarial data augmentation methods for parallel and non-parallel corpora are proposed in Sec. \ref{sec:adv_parallel} and \ref{sec:adv_non_parallel} respectively. 
Sec. \ref{sec:exp} presents the experimental results of using augmented data for training hybrid TDNN and E2E Conformer based ASR systems and analysis. 
Sec. \ref{sec:conclusion} draws the conclusion and discusses possible future works.

\section{Signal Based Dysarthric and Elderly Speech Data Augmentation}
\label{sec:traditional_da}

In this section, we present two traditional front-end signal level data augmentation methods, tempo and speed perturbation for dysarthric and elderly speech recognition. 
These serve as the baseline augmentation approaches to modify the overall speaking rate, volume and spectral shape, and the necessary tempo alignment between normal speed and dysarthric, or elderly speech utterances to facilitate GAN based augmentation model training in Sec. \ref{sec:adv_parallel}.

\subsection{Tempo Perturbation}
\label{subsec:tempo_perturb}
Tempo perturbation modifies the duration of the input time-domain signal $x(t)$, while keeping its overall contour the same \cite{verhelst1993overlap, kanda2013elastic}. This is often implemented using the waveform overlap-add (WSOLA) algorithm \cite{verhelst1993overlap}   that includes three processing stages: signal decomposition, frame relocation and adaptation and signal reconstruction.

After decomposing input audio signal $x(t)$ into analysis blocks $\tilde{x}_m(r)$ that are equally distributed along time axis by analysis hopsize $H_\alpha$, the blocks are relocated based on a method named overlap-add (OLA) \cite{short1977allen}.
OLA relocates analysis blocks $\tilde{y}(r)$ along time axis based on the following equation and a given perturbation factor $\alpha$,
\begin{equation}
	H_s = \alpha \times H_\alpha
\end{equation}
where $H_s$ stands for synthesis hopsize.  

In the next frame relocation stage,
to ensure the perturbed output $y(t)$ has the maximal similarity with $x(t)$, WSOLA 
uses an iterative approach to update the positions of analysis blocks.
For an analysis block $\tilde{x}_m(r)$, its center is shifted by $\Delta_m \in [-\Delta_{max}, \Delta_{max}]$ along the time axis, where the optimal value of $\Delta_{max}$ is obtained by maximizing the cross-correlation between $\tilde{x}_{m}(r)$ and $\tilde{x}_{m-1}(r)$. 
This ensures that the periodic structures of the adjusted analysis block are optimally aligned with the one of the previously copied synthesis block in the overlapping region while both blocks use $H_s$. The Hann window function $w(r)$ is then applied to the adjusted analysis block to compute the synthesis block $\tilde{y}_m(r)$. 

In the final stage, after finishing all iterations, the synthesis frames are processed in order to reconstruct the actual time-scale modified output signal $y(t)$ in a similar manner as conventional OLA \cite{short1977allen}.
The perturbed signal $y(t)$ has a different duration, for example, representing a slower speaking rate of an impaired speaker, but keeps the overall spectral shape the same as $x(t)$.

\subsection{Speed Perturbation}
\label{subsec:speed_perturb}
Speed perturbation \cite{ko2015audio} modifies the input time domain speech signal $x(t)$ by scaling the sampling resolution via a perturbation factor $\alpha$. 
The resulting speed perturbed signal output $y(t)$ is given as: $y(t)=x(\alpha t)$. 
The above time-domain signal modification is equivalent to the following performing in the frequency domain:
\begin{equation}
	X(f) \rightarrow \frac{1}{\alpha}X(\frac{1}{\alpha}f)
\end{equation}
where $X(f)$ and $\frac{1}{\alpha}X(\frac{1}{\alpha}f)$ denote the Fourier transform of $x(t)$ and $y(t)$ respectively.
Speed perturbation changes both audio duration and overall spectral shape~\cite{ko2015audio}. 
This serves to emulate, for example, slower speaking rates, changes in speech volume and formant positions of impaired or elderly speakers.

\begin{figure*}[!t]
	\centering
	\subfloat[]{
		\includegraphics[height=0.23\linewidth]{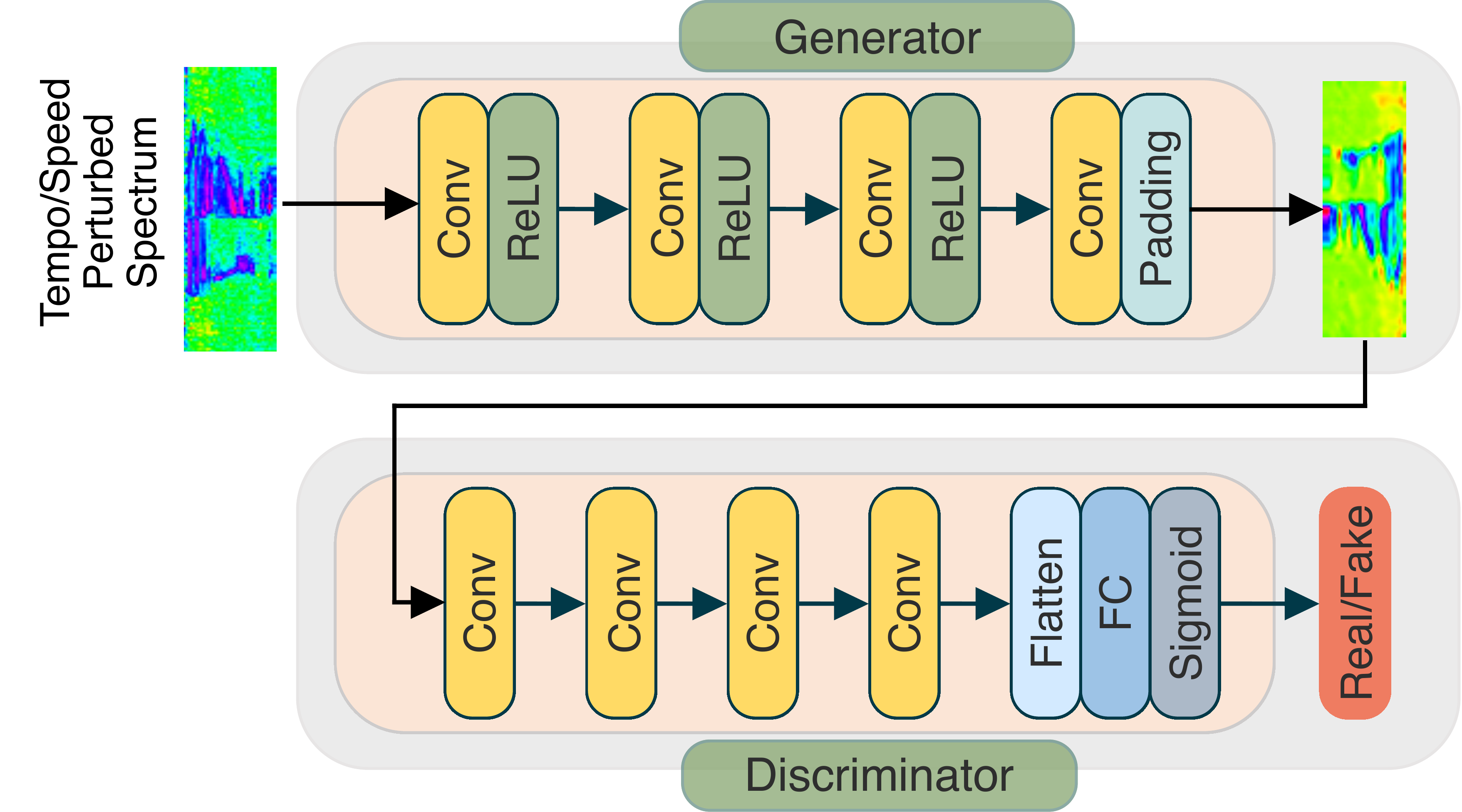}
		\label{subfig:dcgan_architecture}
	}\hfill
	\subfloat[]{
		\includegraphics[height=0.23\linewidth]{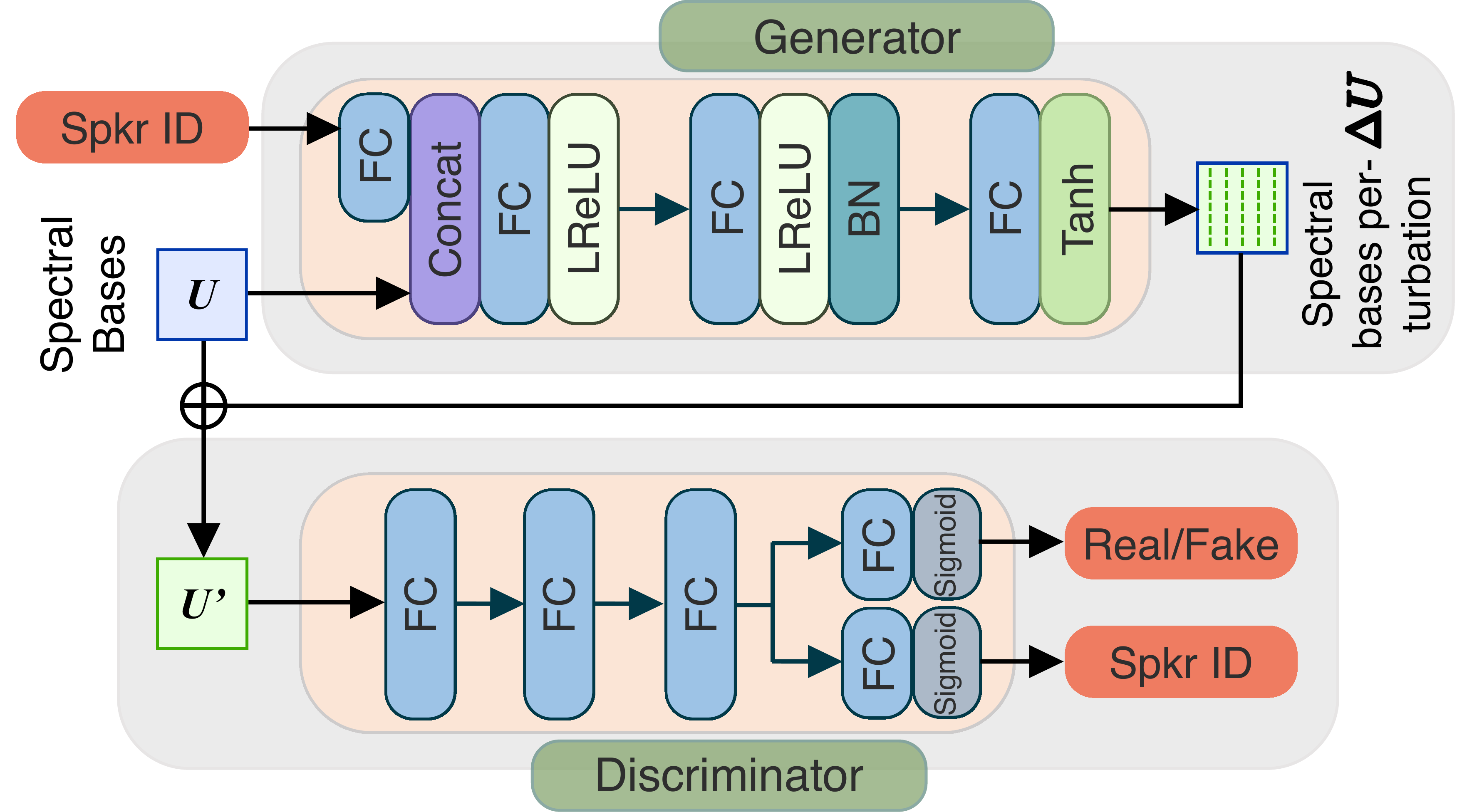}
		\label{subfig:gan_architecture}
	}
\caption{The architectures of the proposed adversarial neural network models designed for impaired or elderly speaker dependent data augmentation using (a) parallel normal and dysarthric speech utterances of identical contents shown in Fig. \ref{fig:subfig:training}-\ref{fig:subfig:conversion}; and (b) non-parallel normal, non-aged and elderly speech data shown in Fig. \ref{fig:subfig:training_non_parallel}-\ref{fig:subfig:conversion_non_parallel}.}
\end{figure*}

\subsection{Speaker Dependent Perturbation Factor Estimation}
\label{subsec:sd_factor}

Speed or tempo perturbation based data augmentation \cite{ko2015audio} are widely used in current ASR systems. 
During the data augmentation process, a group of speaker independent (SI) speed perturbation factors based on, for example, $\{0.9, 1.0, 1.1\}$, are applied at the front-end level to produce a three-fold expansion of the original training data. 
Prior researches further suggest that additional use of speaker dependent (SD) perturbation of normal, healthy speech to expand the limited training data for each target dysarthric \cite{geng2021investigation, liu2021recent,vachhani2018data, xiong2019phonetic} or elderly speaker \cite{ye2021development} during data augmentation produced better ASR performances over using the speaker independent perturbation approach only. 
More detailed analyses on such combined use of speaker independent and dependent speed perturbation for dysarthric speech recognition can be found in \cite{geng2021investigation, liu2021recent}.

Without the loss of genericity, for any target dysarthric or elderly speaker, the associated SD perturbation factor is calculated as the average phoneme duration ratio between their respective speech obtained using phoneme alignment analysis \cite{xiong2019phonetic}. 
Force alignment using a HTK~\cite{young2002htk} trained GMM-HMM system is first performed. 
The resulting frame-level phoneme alignments are then used to compute the SD perturbation factor $\alpha_{j}$ for the $j^{th}$ target dysarthric or elderly speaker as $\alpha_j = \frac{\bar  d_{C}}{d_j}$, where $\bar{d_C}$ denotes the average time duration of all healthy, control speakers and $d_j$ is the average phoneme duration of the target $j^{th}$ dysarthric or elderly speaker. 

\section{Adversarial Augmentation Using Parallel Data}
\label{sec:adv_parallel}

As discussed in Sec. \ref{sec:intro}, conventional data augmentation methods based on tempo or speed perturbation in Sec. \ref{sec:traditional_da} only characterize an overall decrease in speaking rate, speech volume and changes in spectral envelope of dysarthric and elderly speech. 
In this section, for dysarthric speech corpora where the underlying spoken contents are designed to be parallel but the speech duration is allowed to vary, speaker dependent deep convolutional GANs (DCGANs) are utilized to learn the relation between tempo or speed perturbed healthy speech spectra and those of individual target impaired speakers. 

\begin{figure}[!t]
    \centering
  	\subfloat[control]{
		\includegraphics[height=1.35cm]{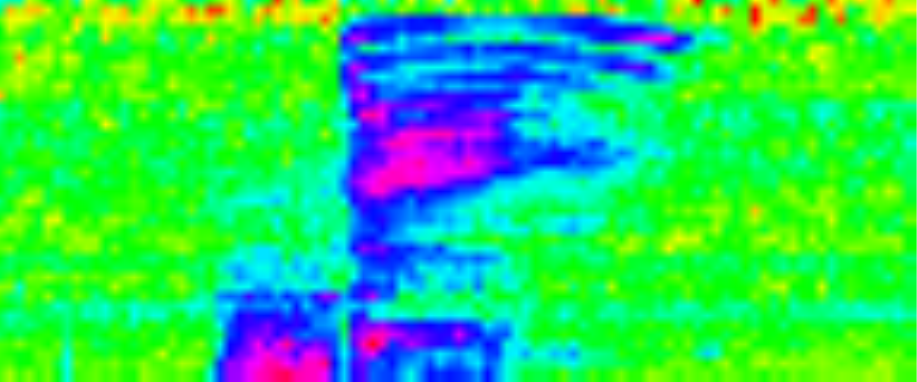}
	    \label{fig:subfig:ctrl}
	  }\hfill
	  \subfloat[dysarthric]{
		\includegraphics[height=1.35cm]{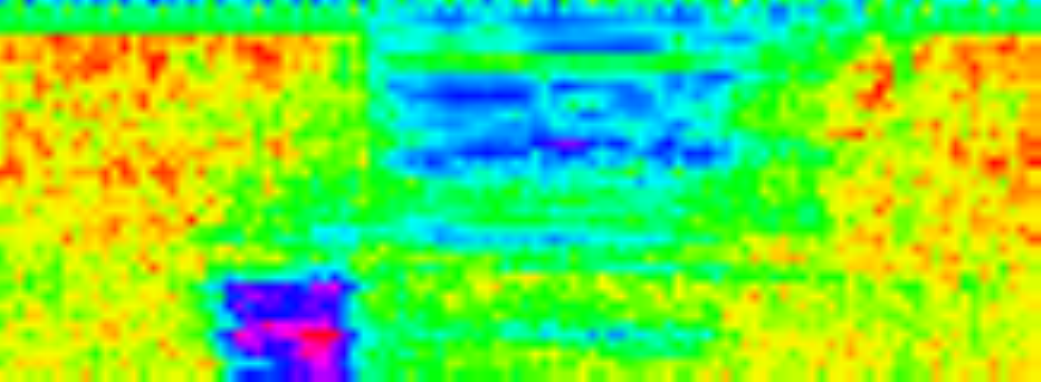}
	    \label{fig:subfig:dys}
	  }
	  
	  \subfloat[tempo]{
    	\includegraphics[height=1.35cm]{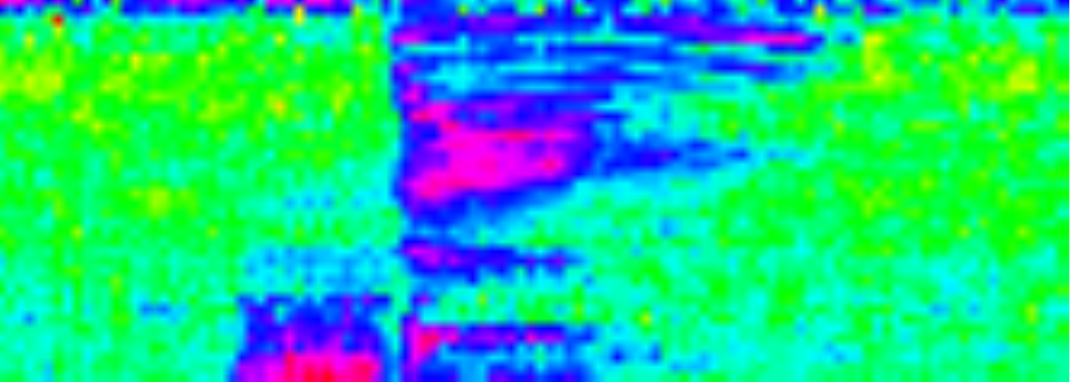}
	    \label{fig:subfig:tempo_ctrl}
	  }\hfill
	  \subfloat[speed]{
    	\includegraphics[height=1.35cm]{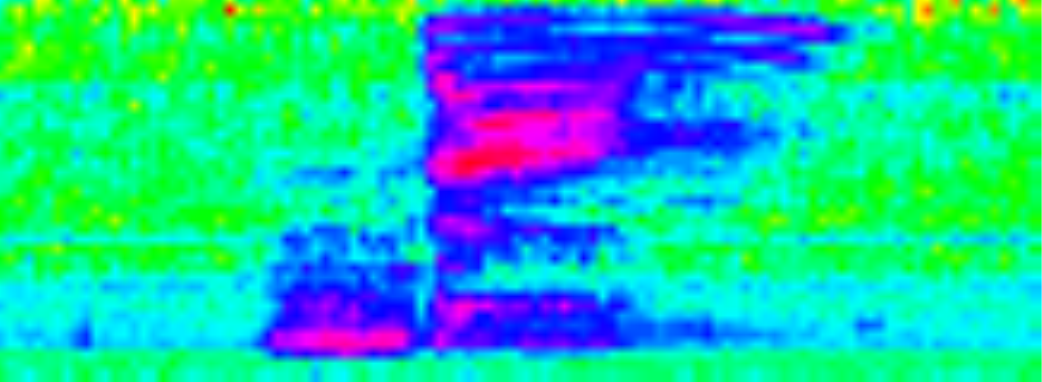}
	    \label{fig:subfig:speed_ctrl}
	  }

	  \subfloat[tempo-GAN]{
    	\includegraphics[height=1.35cm]{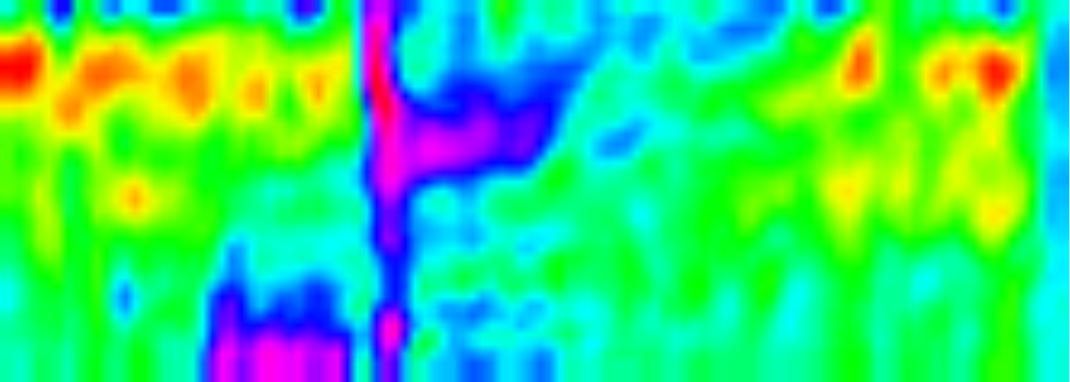}
	    \label{fig:subfig:tempo_gan}
	  }\hfill
	  \subfloat[speed-GAN]{
    	\includegraphics[height=1.35cm]{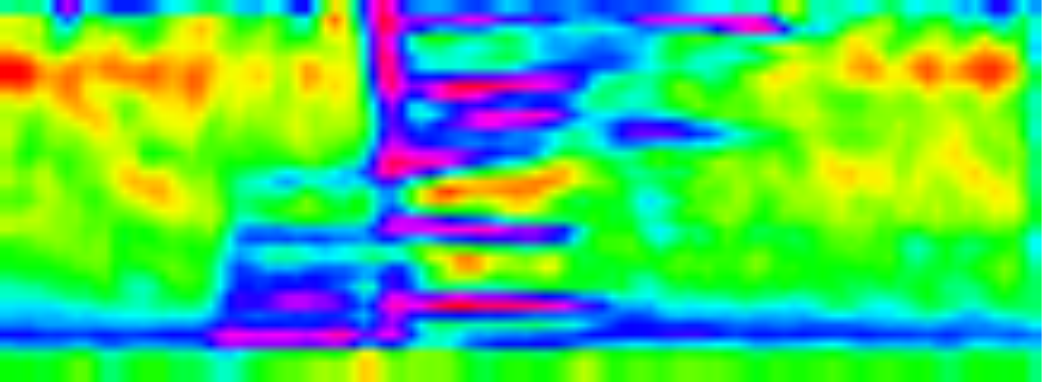}
	    \label{fig:subfig:speed_gan}
	  }
  \caption{Example spectrogram of (a) control, (b) dysarthric, (c) tempo or (d) speed perturbed control speech, and (e) tempo-GAN or (f) speed-GAN generated speech.}
  \label{fig:ctrl_and_dys_fbank}
\end{figure}

\subsection{DCGAN Model Architecture}
\label{subsec:parallel_gan_arch}
The overall architecture configurations of the proposed DCGAN model follow  our previous work \cite{jin21_interspeech}, also again shown in Fig. \ref{subfig:dcgan_architecture}.
The Generator component contains 4 convolutional layers, the first three of which have 8 kernels while the last one has 1 kernel only. 
All of these kernels in the Generator have a kernel size of $3 \times 3$ and stride of $1 \times 1$. 
Each of the first three convolutional layers is also immediately connected to ReLU activations. 
We use Replicate Padding to replicate the edges of the feature map to ensure the output and input dimensions are the same.
The Discriminator component contains 4 convolutional layers of 8, 16, 32 and 64 kernels respectively, all of which use a kernel size of $2 \times 2$ and stride of $2 \times 2$. 
A flattening operation is applied to concatenate the outputs of convolutional layers, resulting in a 3000-dimensional vector.
A fully connected (FC) layer with Sigmoid activation is used for binary classification in the Discriminator.

\begin{figure*}[!t]
	\centering
	\subfloat[] {
		\includegraphics[width=0.45\linewidth]{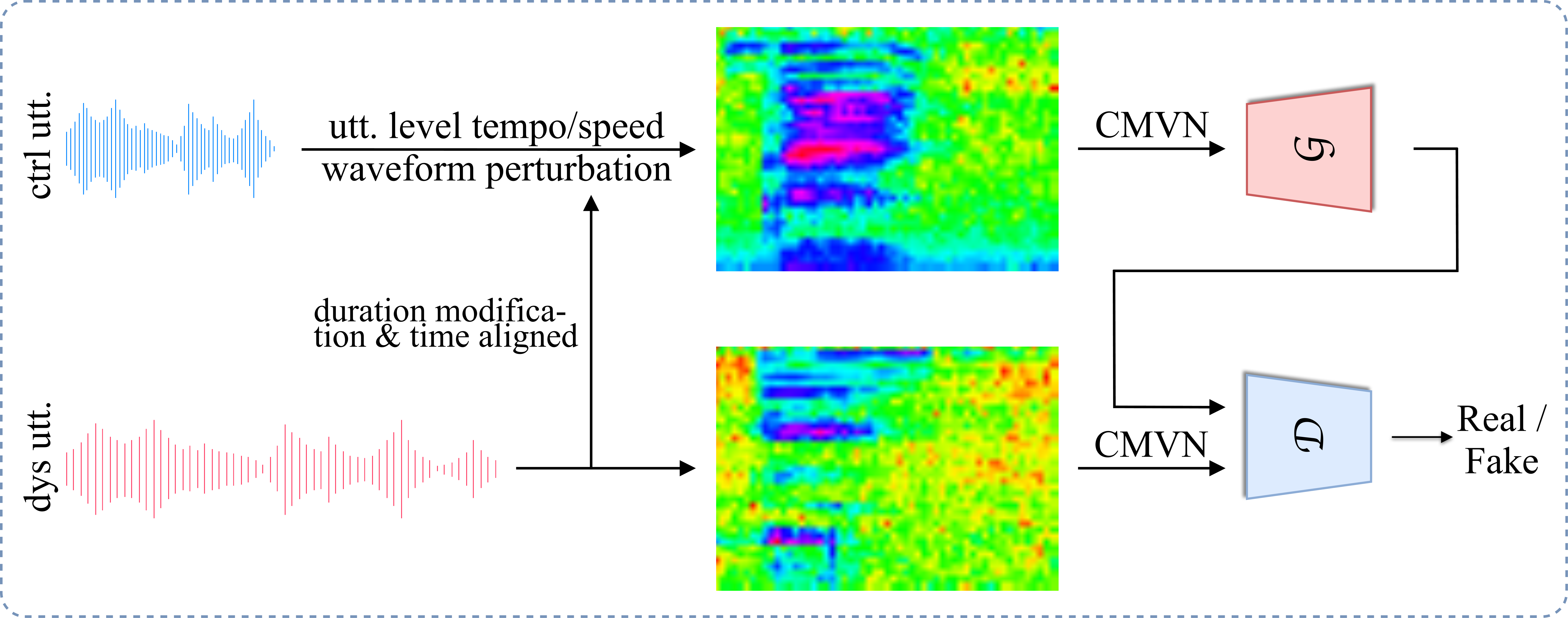}
		\label{fig:subfig:training}
	}\hfill
	\subfloat[] {
		\includegraphics[width=0.45\linewidth]{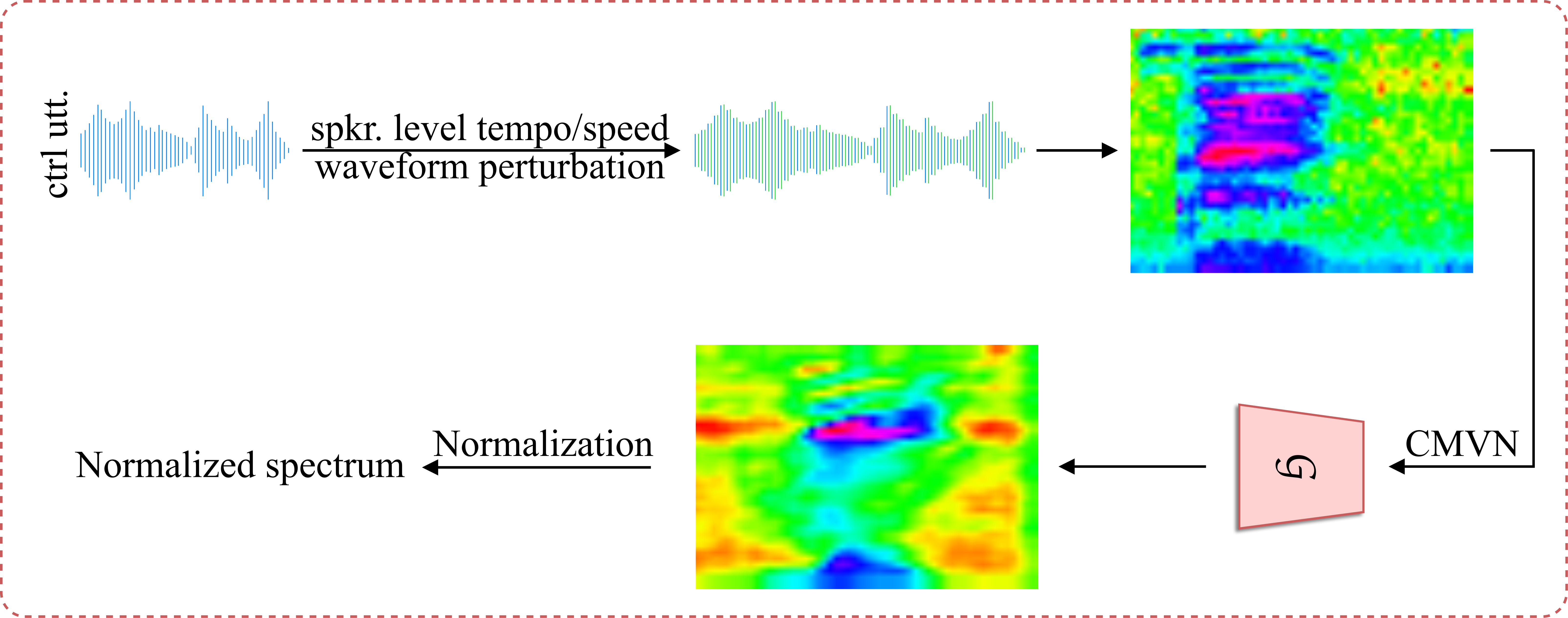}
		\label{fig:subfig:conversion}
	}
	
	\subfloat[] {
		\includegraphics[width=0.45\linewidth]{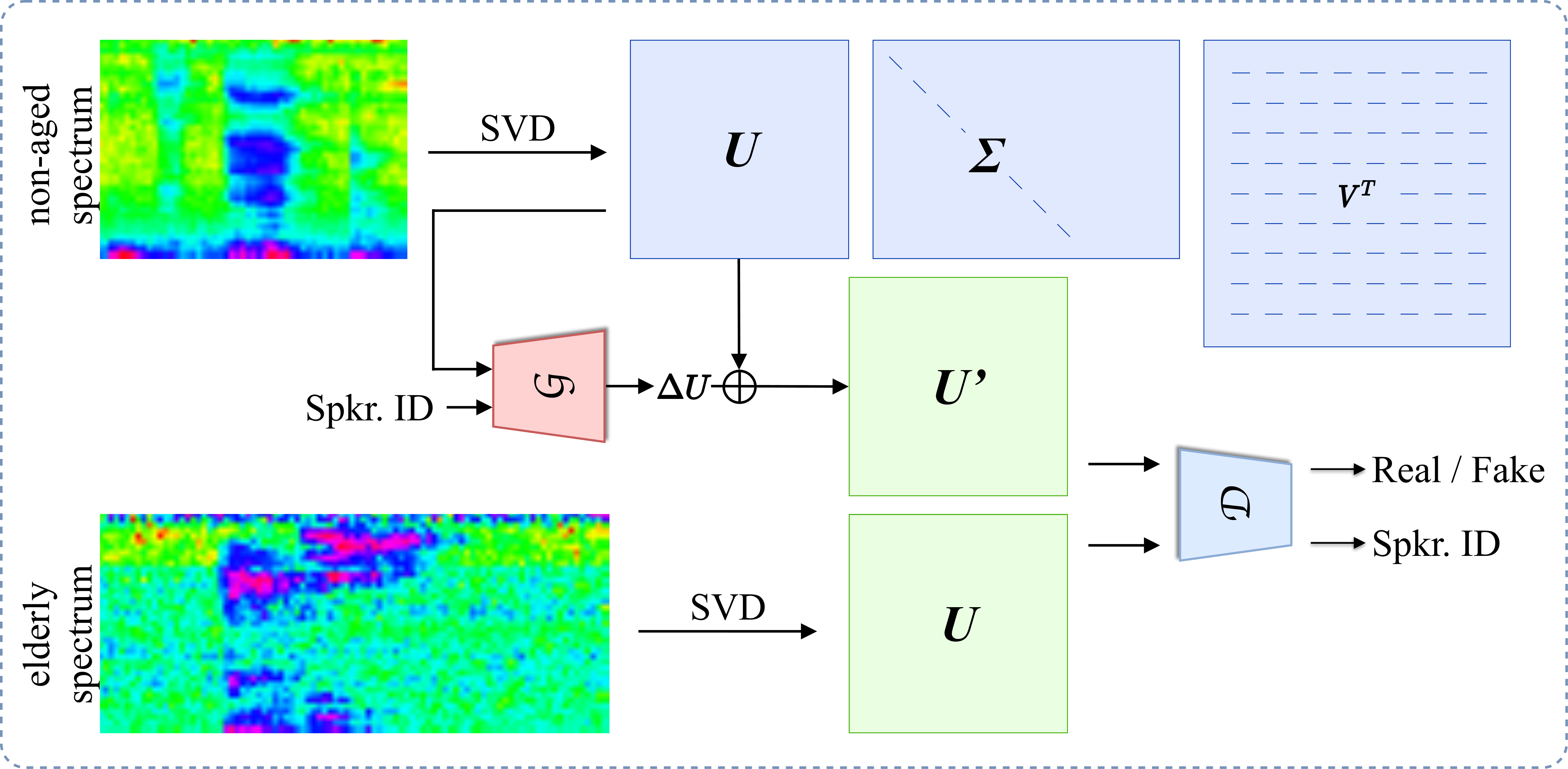}
		\label{fig:subfig:training_non_parallel}
	}\hfill
	\subfloat[] {
		\includegraphics[width=0.45\linewidth]{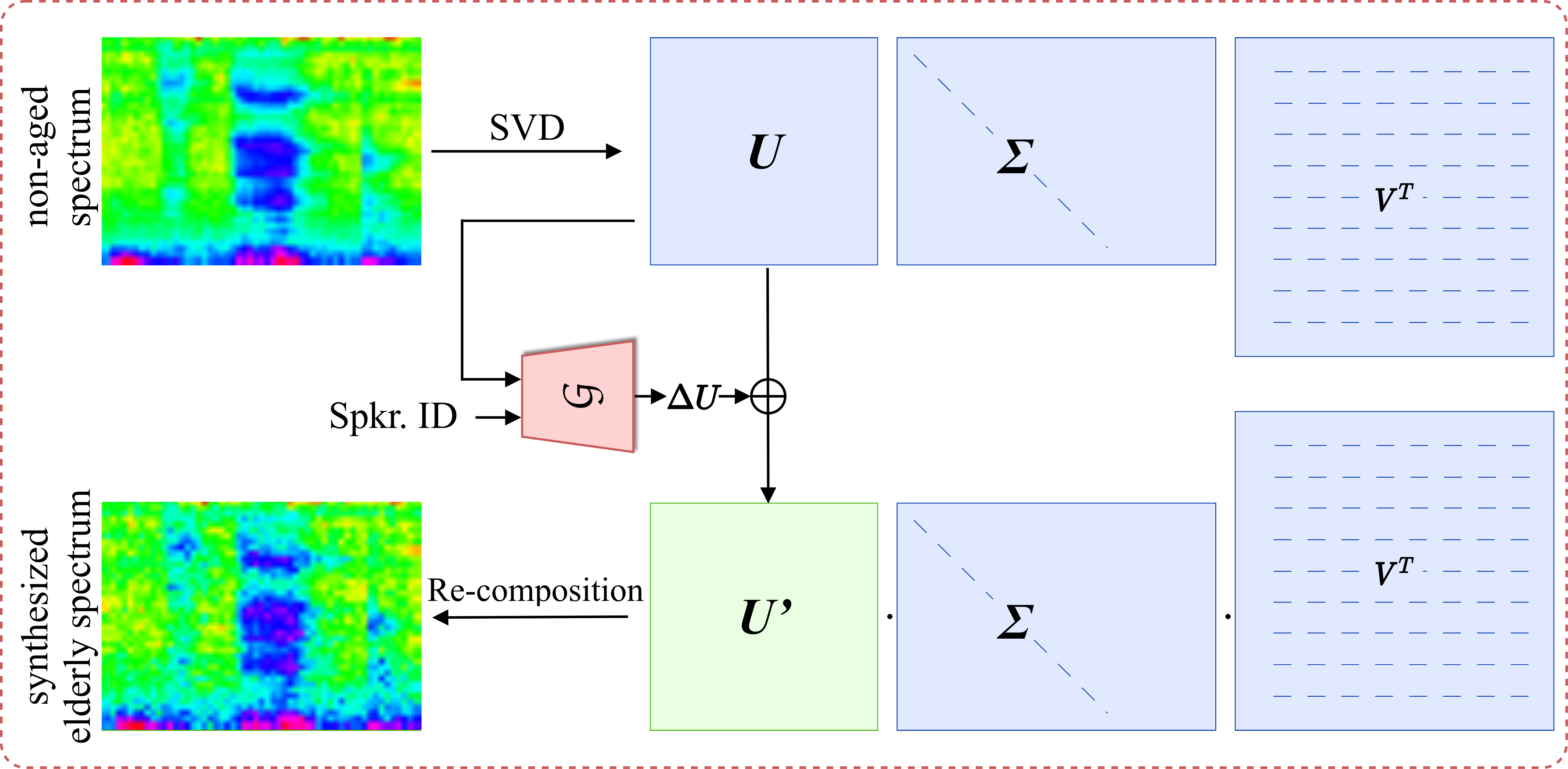}
		\label{fig:subfig:conversion_non_parallel}
	}
	
	\caption{Illustration of (a) DCGAN model training on parallel control and dysarthric utterances with modified duration and time alignment; (b) DCGAN based speaker dependent dysarthric speech spectrogram generation using impaired speaker level tempo/speed perturbed normal speech as the input; (c) Spectral basis GAN model training on SVD decomposed non-parallel non-aged and elderly speech spectrograms; and (d) Spectral basis GAN based speaker dependent elderly speech spectrogram generation by re-composition of perturbed non-aged speech derived spectral basis vectors with their temporal bases. }
	\label{fig:training_non_parallel}
	
  \end{figure*}

\subsection{DCGAN Model Training}
\label{subsec:parallel_gan_training}

Prior to DCGAN training, pairs of normal and dysarthric speech utterances of identical word contents but often different duration need to be formed. 
In order to facilitate a frame-by-frame comparison between the GAN transformed normal speech spectrogram against that of the target impaired speech, each normal speech segment is either tempo or speed perturbed to produce a modified duration that matches against that of a target dysarthric speech utterance, as is shown in Fig. \ref{fig:subfig:training}. 
This requires a scaling factor to be estimated for each normal and dysarthric speech segment pair using phonetic analysis similar to the procedure described in Sec. \ref{subsec:sd_factor} for speaker level speed or tempo perturbation. 
The resulting pairs of normal and dysarthric speech utterances that now have the same duration are further zero mean and unit variance normalized at speaker level, before being used in DCGAN training. 

The DCGAN training objective function both maximizes the binary classification accuracy on the target dysarthric speech spectrum and minimizes that obtained on the GAN transformed normal speech.
This is given by
\begin{equation}
	\begin{aligned}
		\mathop{min}\limits_{G_j}^{}\mathop{max}\limits_{D_j}^{} &\ V(D_j, G_j) \\
		&\ = \mathbb{E}_{\textbf{f}_{\textbf{D}}\sim p_{D_j}(\textbf{f})}[\log{(D_j(\textbf{f}_{\textbf{D}_\textbf{j}}))}] \\
		&\ + \mathbb{E}_{\textbf{f}_{\textbf{C}} \sim p_{C}(\textbf{f})}[\log{(1-D_j(G_j(\textbf{f}_{\textbf{C}})))}]
	\end{aligned}
\end{equation}
where $j$ represents the index for target dysarthric speaker, $G_j$ and $D_j$ are Generator and Discriminator associate with dysarthric speaker $j$, $\textbf{f}_\textbf{C}$ and $\textbf{f}_{\textbf{D}_\textbf{j}}$ stand for the Mel-scale filter-bank (FBank) features of paired control and dysarthric utterances.
During DCGAN training for each target impaired speaker, the learning rate for both the Generator and Discriminator is halved every 2500 iterations until convergence.

\subsection{Dysarthric Speech Spectrum Generation}
\label{subsec:parallel_gan_generation}

During DCGAN based data augmentation, as shown in Fig. \ref{fig:subfig:conversion}, the target impaired speaker level tempo or speed perturbed 40-dimensional Mel-scale filter-bank features obtained from normal speech using the baseline tempo or speed perturbation approaches described in Sec. \ref{sec:traditional_da} are fed into the corresponding Generator components to produce the comparable speaker level ``tempo-GAN'' or ``speed-GAN'' augmented data for each target impaired speaker, before being further zero mean and unit variance normalized for ASR system training. 

The resulting data augmentation process not only allows an overall change in speaking rate, speech volume and spectral shape to be produced as in conventional signal level tempo or speed perturbation, but also injects further detailed spectro-temporal characteristics associated with impaired speech including articulatory imprecision, decreased clarity as well as breathy and hoarse voice into the final augmented filter-bank data for each target impaired speaker. 

For instance, in contrast to an example control speech segment's spectrogram of the word ``Some'' in Fig. \ref{fig:subfig:ctrl}, the comparable dysarthric speech spectrogram in Fig. \ref{fig:subfig:dys} contains not only weakened formants that indicates articulatory imprecision, but also additional energy distributed over higher frequency components at both the start and end of the utterance due to the difficulty in breath control when speaking. 
Such additional energy is more clearly captured in the speed-GAN generated spectrogram shown in Fig. \ref{fig:subfig:speed_gan}, than that derived using speed perturbation only shown in Fig. \ref{fig:subfig:speed_ctrl}. 

\section{Adversarial Augmentation Using Non-Parallel Data}
\label{sec:adv_non_parallel}

The adversarial data augmentation approach introduced in Sec. \ref{sec:adv_parallel} requires the use of parallel control and dysarthric or elderly speech recordings of identical spoken contents. 
In this section, SVD based speech spectrum decomposition is used to derive spectral and temporal subspace representations. 
Among these, the spectral basis features of a healthy, non-aged source speaker considered to be more closely related to time invariant speaker characteristics, are transformed via SD spectral basis perturbation GANs into those of a target elderly or dysarthric speaker, before being re-composed with the normal speech temporal bases to produce the augmented data.  

\begin{CJK}{UTF8}{gbsn}
	\begin{figure*}[!ht]
		\centering
		{
			\includegraphics[width=0.45\linewidth]{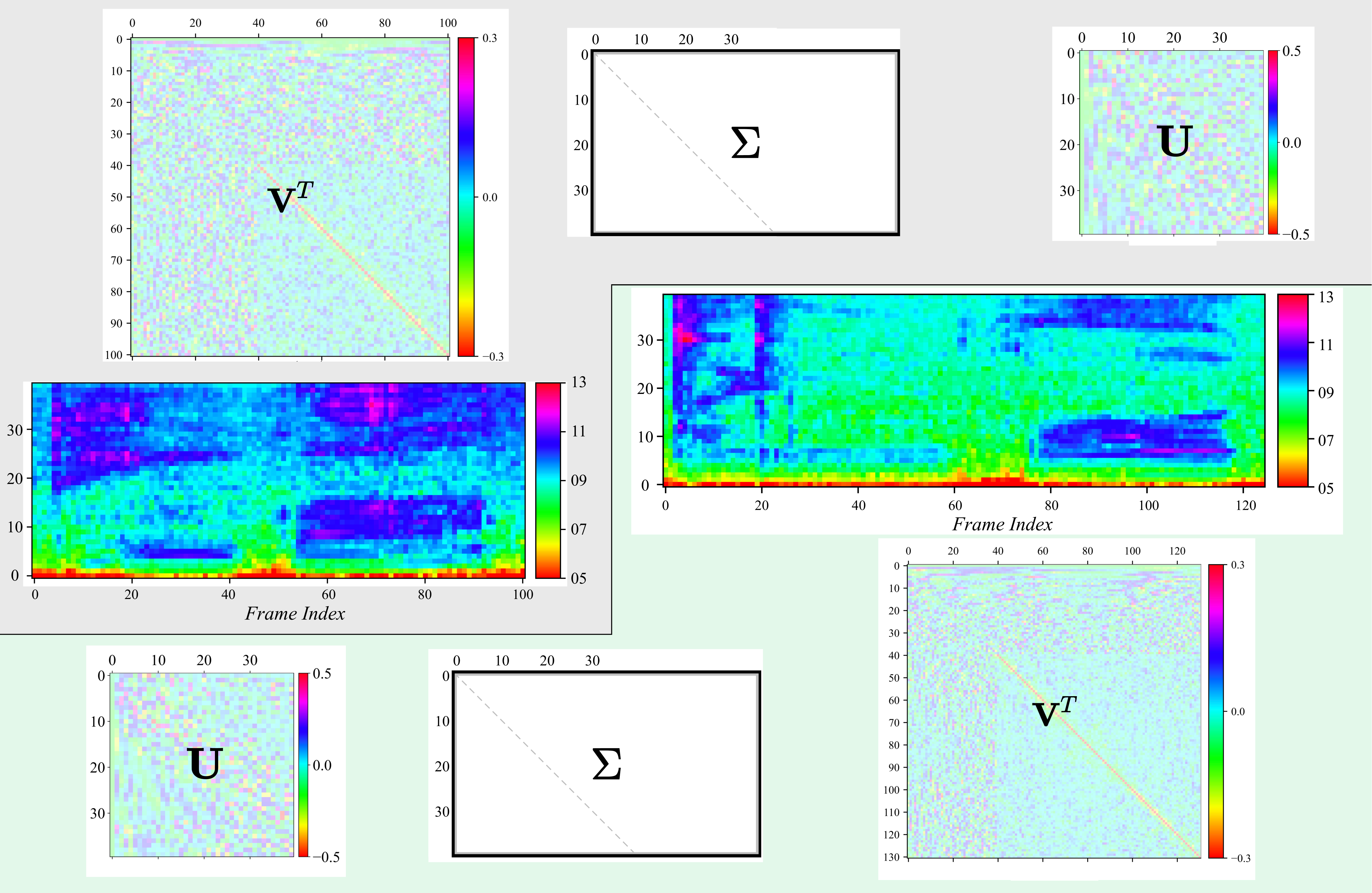}
		}\hfill
		{
			\includegraphics[width=0.45\linewidth]{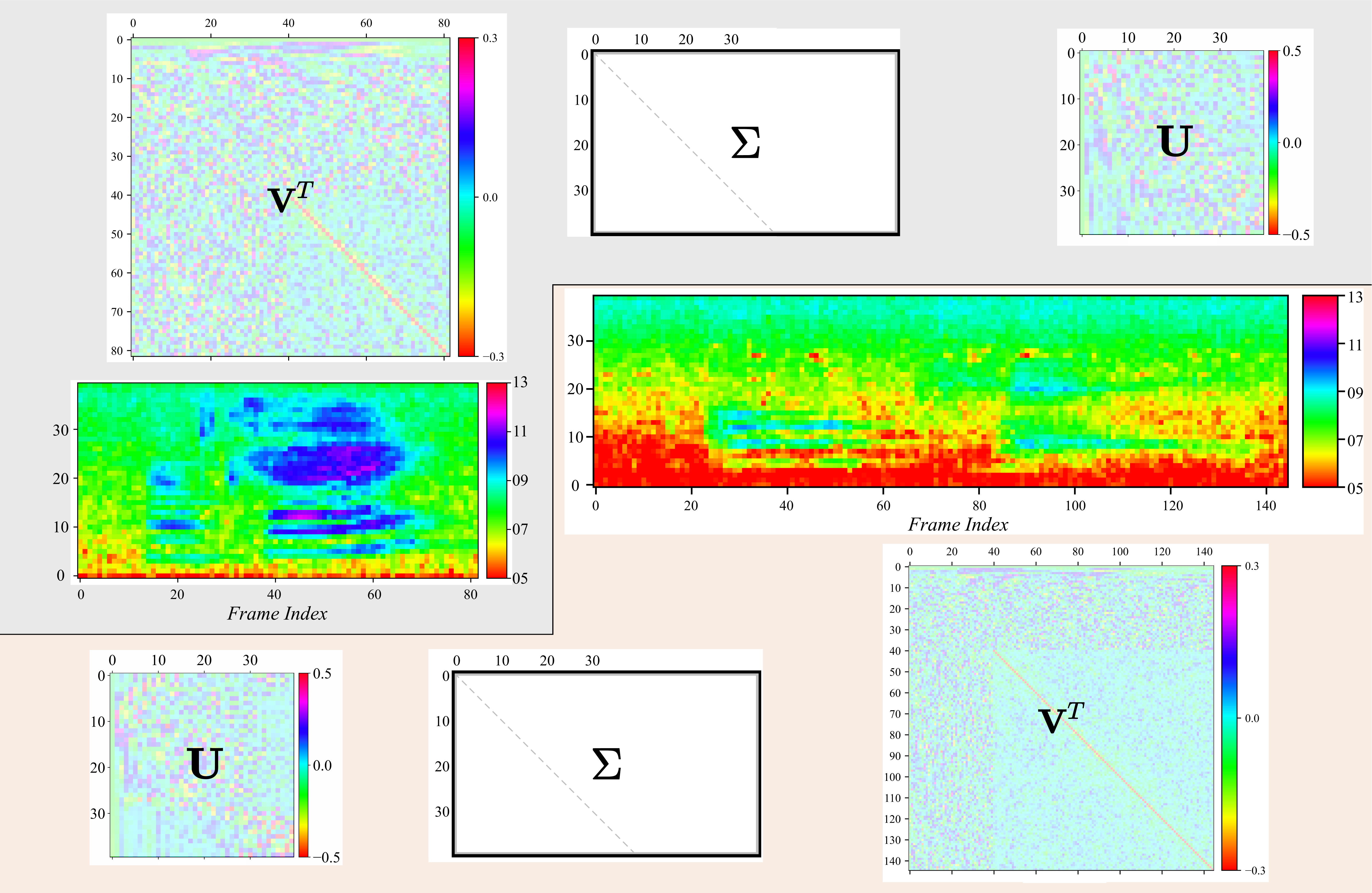}
		}
	\caption{Example subspace decomposition of Mel-spectrograms of: (a)  a pair of elderly participant (PAR, right middle in (a)) and non-aged clinical investigator (INV, left middle in (a)) utterances of the Cantonese word ``苹果 (apple)'' produced spectral and temporal basis vectors ($\mathbf{U}$ and $\mathbf{V^{\mathrm{T}}}$) of the JCCOCC MoCA (JCMOCA) \cite{xu2021speaker} corpus; and (b) a pair of elderly participant (PAR, left middle in (b)) and non-aged clinical investigator (INV, right middle in (b)) utterances of the English word ``okay'' produced spectral and temporal basis vectors ($\mathbf{U}$ and $\mathbf{V^{\mathrm{T}}}$) of the DementiaBank Pitt (DBANK) \cite{becker1994natural} dataset.}
	\label{fig:svd_example}
	\end{figure*}
	\end{CJK}

\subsection{Speech Spectrum Subspace Decomposition}
\label{subsec:speech_svd}

Spectro-temporal subspace spectrum decomposition techniques provide simple and intuitive solutions to decouple the time invariant spectral components of speech signals from their time variant temporal components by modelling the combination between these two using a linear system \cite{van1993subspace}. 
This linear system can then be solved using signal subspace decomposition schemes, for example, singular value decomposition (SVD)\cite{van1993subspace,janbakhshi2020subspace} or non-negative matrix factorization (NMF) methods \cite{lee1999learning, fevotte2009nonnegative,wang2012online}, both of which are performed on the time-frequency speech spectrum. 

Let $\textbf{S}_r$ represents a $C \times T$ dimensional Mel-scale spectrogram of utterance $r$ with $C$ filter-bank channels and $T$ frames.
The SVD decomposition \cite{van1993subspace} is given by:
\begin{equation}
	\textbf{S}_r = \textbf{U}_r \vb{\Sigma}_r \textbf{V}^T_r
\end{equation}
where the set of column vectors of the $C \times C$ dimensional left singular $\textbf{U}_r$ and the row vectors of the $T \times T$ dimensional right singular $\textbf{V}^T_r$ stands for time invariant spectral basis vector and time variant temporal basis vector respectively.
Here $\vb{\Sigma}_r$ is a $C \times T$ rectangular diagonal matrix containing the singular values sorted in descending order. 

\begin{CJK}{UTF8}{gbsn}
Two examples of SVD decomposition of Mel-scale filter-bank based log amplitude spectra associated with (a) an Cantonese speech utterance containing the same word ``苹果 (apple)'' of the JCCOCC MoCA corpus \cite{xu2021speaker}, and (b) another English pair containing the word ``okay'' of the DementiaBank Pitt \cite{becker1994natural} dataset, are shown in Fig. \ref{fig:svd_example} and \ref{fig:svd_example} respectively. 
The resulting spectral and temporal basis matrices intuitively represent the following two sources of information:
\textbf{1) time-invariant spectral subspaces: } 
that can be associated with an average utterance-level description of elderly speakers' characteristics such as an overall reduction of speech volume, 
weakened formants due to articulation imprecision as well as hoarseness and energy distribution anomaly across frequencies due to difficulty in breath control. 
For example, the comparison between the spectral basis vectors extracted from a pair of Cantonese elderly and normal speech utterances of the same word ``苹果 (apple)'' in Fig. \ref{fig:svd_example} (bottom, left) shows that the elderly spectral basis matrix exhibits a sparser 
energy distribution pattern 
than those obtained from the comparable normal, non-aged speech spectral bases in Fig. \ref{fig:svd_example} (top, right). 
Similar trends can be found between the spectral basis vectors of the non-aged and elderly speech utterances of the same English word content ``okay'' shown in Fig. \ref{fig:svd_example}. and 
\textbf{2) time-variant temporal subspaces: } 
that are considered more related to time content dependent features such as duration and pauses, for example, shown in the contrast between the temporal basis vectors separately extracted from non-aged and elderly speech in Fig. \ref{fig:svd_example} and Fig. \ref{fig:svd_example}. 
The dimensionality of the temporal bases captures the duration. 
\end{CJK}

\subsection{Spectral Basis GAN Model Architecture}
\label{subsec:non_parallel_gan_arch}

The overall architecture of the proposed spectral basis GAN model is shown in Fig. 1b. Compared with the DCGAN of Sec. \ref{sec:adv_parallel} designed for parallel data shown in Fig. \ref{subfig:dcgan_architecture}, several changes are made. 
First, SVD decomposed spectral bases are now used as the GAN inputs instead of filter-bank features. 
Hence, the convolutional layers adopted in the DCGAN model of Fig. \ref{subfig:dcgan_architecture} tailored for filter-bank input features are replaced by fully connected (FC) layers.
More specifically, the Generator (Fig. \ref{subfig:gan_architecture}, top) contains three fully connected layers of $512$, $512$ and $1600$ dimensions. The first two FC layers are both followed by a Leaky ReLU function with a negative slope value set to $0.2$, while the third FC layer is followed by a Tanh activation function before producing the final outputs. 
The Discriminator (Fig. \ref{subfig:gan_architecture}, bottom) also contains 3 FC layers of $256$, $512$ and $256$ dimensions each.  

Second, the output targets of the GAN model serve as a perturbation vector, $\Delta \mathbf{U}$, to be added to the spectral bases that are derived from the input normal, non-aged speech spectrogram, $\mathbf{U}$. 
During GAN training, many-to-many mapping exists between multiple normal, non-aged speech utterances and a given set of target elderly speech segments. 
To this end, three forms of pairing between these two groups of data are considered in the training stage. 
The first adopts a random pairing between a normal, non-aged speech utterance and any randomly selected elderly speech segment of a target speaker. 
The second utilizes a pairing between a normal speech utterance and the average, mean spectral bases computed over all the elderly speech utterances of a target speaker. 
The third approach considers a more expensive and exhaustive, full permutation over all possible pairing between a normal speech utterance and each elderly speech utterance of a target speaker. 
In practice, the three normal and elderly speech data pairing strategies produced comparable performance, as will be shown later in the experimental results of Sec. \ref{sec:exp}.

Finally, for the speaker dependent spectral basis GAN model considered here, it is vital to ensure speaker level homogeneity, for example, an overall reduced speech volume, to be consistently encoded in resulting spectral basis perturbation output vectors. 
To this end, in addition to the binary non-aged versus elderly classification task, a second auxiliary task based on elderly speaker ID prediction is also used in the Discriminator training, akin to the Auxiliary Classifier GAN (AC-GAN) introduced in \cite{pmlr-v70-odena17a}.
As shown in the bottom right corner of Fig. \ref{subfig:gan_architecture}, for each of these two tasks, a separate 256-dimensional FC layer followed by a Sigmoid activation is used. 
Additional speaker ID one-hot encoding features are fed into the Generator as shown in the top left corner of Fig. \ref{subfig:gan_architecture}.

\subsection{Spectral Basis GAN Model Training}
\label{subsec:non_parallel_gan_training}

An example illustration of spectral basis GAN model training is shown in Fig. \ref{fig:subfig:training_non_parallel} on a pair of spectral basis matrices separately derived from SVD decomposition of non-parallel non-aged (top left, $\textbf{U}$ in light blue, Fig. \ref{fig:subfig:training_non_parallel}) and elderly (bottom left, $\textbf{U}$ in light green, Fig. \ref{fig:subfig:training_non_parallel}) speech spectrograms. 
The GAN Generator synthesized elderly speech spectral basis matrix, $\textbf{U'} = \textbf{U} + \Delta \textbf{U}$, adds the GAN output perturbation vector to the non-aged speech spectral bases (centre, Fig. \ref{fig:subfig:training_non_parallel}), before being fed into the Discriminator to perform binary classification over the non-aged versus aged labels and speaker ID prediction in the second auxiliary task introduced above in Sec. \ref{subsec:non_parallel_gan_arch} and Fig. \ref{subfig:gan_architecture}. 
The spectral basis GAN model training loss function is given by:
\begin{equation}
	\begin{aligned}
		\mathop{min}\limits_{G_j}^{} &\ \mathop{max}\limits_{D_j}^{} V(D_j, G_j) \\
		&\ = \mathbb{E}_{\textbf{U}_{\textbf{D}}\sim p_{j}(\textbf{U})}[\log{(D_j(\textbf{U}_{j}))}] \\
		&\ + \mathbb{E}_{\textbf{U}_{\textbf{C}} \sim p_{C}(\textbf{U})}[\log{(1-D_j(\textbf{U}_\textbf{C} + \Delta \textbf{U}))}]
	\end{aligned}
\end{equation}
where $j$ , $G_j$ and $D_j$ are the Generator and Discriminator modules associated with a target elderly speaker $j$, $\textbf{U}_\textbf{C}$ and $\textbf{U}_j$ denote the spectral bases that are separately derived from SVD decomposition of non-parallel non-aged and elderly speech utterance spectrograms. 
$\Delta \textbf{U} = \lambda G_{j}(\textbf{U}_\textbf{C}) $ represents the generated spectral bases perturbation vector.
$\lambda$ is a scaling factor used to control and moderate the perturbation added to the non-aged speech spectral basis vectors, and empirically set as $0.1$ and $0.2$ for dysarthric and elderly speech datasets respectively.
Further ablation study on the setting of $\lambda$ and its impact on ASR system performance is provided later in the experiment section of this paper (in Tab. \ref{tab:lambda_dbank}). 

For the two output tasks, their respective error cost functions are:
\begin{equation}
	\label{eq:loss_c}
	\begin{aligned}
		L_c = &\ \mathbb{E}[\log{P(Cond.=\textbf{E} | \textbf{U}_{j})}] \\
		&\ + \mathbb{E}[\log{P(Cond.=\textbf{C} | \textbf{U}_\textbf{C} + \Delta \textbf{U})}]
	\end{aligned}
\end{equation}
\begin{equation}
	\label{eq:loss_s}
	\begin{aligned}
		L_{sid} = &\ \mathbb{E}[\log{P(Sid=j | \textbf{U}_{j})}] \\
		&\ + \mathbb{E}[\log{P(Sid=j | \textbf{U}_\textbf{C} + \Delta \textbf{U})}]
	\end{aligned}
\end{equation}
where $\textbf{U}_j$ stands for the spectral bases extracted from the spectrogram of an utterance of a target elderly or dysarthric speaker $j$.
The above classification cost $L_c$ is formulated such that $Cond. = E$ and $Cond. = C$ state whether the real spectral bases, $\textbf{U}_j$, or a synthesized one, $\textbf{U}_C + \Delta \textbf{U}$, are produced by an elderly speaker or a control, non-aged speaker respectively.
The speaker ID prediction cost $L_{sid}$ ensures both the original and synthesized spectral bases share the same target speaker specific characteristics.  
The Discriminator is trained to maximize $L_{sid} + L_c$ while the Generator is trained to maximize $L_{sid} - L_c$. 
For both Generator and Discriminator components, the learning rate is halved for every $2500$ iterations until convergence.

\subsection{Elderly Speech Spectrum Generation}
\label{subsec:non_parallel_gan_generation}

An example illustration of spectral basis GAN based speaker dependent elderly speech spectrogram generation is shown in Fig. \ref{fig:subfig:conversion_non_parallel}. 
This is implemented by re-composition of the target speaker dependent GAN perturbed spectral basis vectors derived from a non-aged speech utterance of arbitrary word contents with its corresponding original temporal bases and singular value matrix. 
This data augmentation process is applied to both the English DementiaBank Pitt \cite{becker1994natural} and Cantonese JCCOCC MoCA \cite{xu2021speaker} elderly speech datasets, both of which only provide non-parallel non-aged investigator and elderly participant speech recordings.

\section{Experiments}
\label{sec:exp}

\begin{table*}[!t]
	\centering
	\caption{Performance of GAN data augmentation approaches on various sizes of expanded training data before and after LHUC-SAT speaker adaptation on the \textbf{UASpeech} \cite{kim2008dysarthric} test set of 16 dysarthric speakers. ``$6$M'' and ``$19$M'' refer to the number of model parameters. ``CTRL'' in the ``Data Augmentation'' column stand for dysarthric speaker dependent transformation of control speech during data augmentation using from left to right: temporal or speed perturbation in ``T'' and ``S''; ``TG'' and ``SG'' denote the tempo-GAN and speed-GAN models of Sec. \ref{sec:adv_parallel}; ``SBG'' stands for the spectral basis GAN of Sec. \ref{sec:adv_non_parallel}; and ``SBG+SG'' denotes spectral basis GAN perturbed normal speech being further transformed using speed-GAN. ``VL/L/M/H'' refers to intelligibility subgroups (Very Low / Low / Medium / High). ``DYS'' column denotes speaker independent speed perturbation of dysarthric speech. $\dag$ and $\star$ denote statistically significant improvements ($\alpha=0.05$) are obtained over the comparable baseline systems with tempo perturbation (Sys. 2, 8, 14, 21) or speed perturbation (Sys. 3, 9, 15, 22) respectively.}
	\label{tab:uaspeech}
    \renewcommand\tabcolsep{3.0pt}
	\scalebox{0.85}{
	\begin{tabular}{c|c|c|c|c|c|c|c|c|c|c|c|c|c|c||c|c|c|c|c}
	    \hline
	    \hline
		\multirow{3}{*}{Sys.} & \multirow{3}{*}{\makecell[c]{Model\\(\# Parameters)}} & \multicolumn{7}{c|}{Data Augmentation} & \multirow{3}{*}{\# Hrs.} & \multicolumn{5}{c||}{WER \% (Unadapted)} & \multicolumn{5}{c}{WER \% (LHUC-SAT Adapted)} \\
		\cline{3-9}
		\cline{11-20}
		& & \multicolumn{6}{c|}{CTRL} & DYS &  & \multirow{2}{*}{VL} & \multirow{2}{*}{L} & \multirow{2}{*}{M} & \multirow{2}{*}{H} & \multirow{2}{*}{Avg.} & \multirow{2}{*}{VL} & \multirow{2}{*}{L} & \multirow{2}{*}{M} & \multirow{2}{*}{H} & \multirow{2}{*}{Avg.} \\
		\cline{3-9}
		& & T & S & TG & SG & SBG & SBG+SG & S &  &  &  &  &  &  &  & &  &  &  \\
		\hline
		\hline
		1 & \multirow{19}{*}{\makecell[c]{Hybrid\\TDNN\\(6M)}} & \multicolumn{6}{c|}{-} & -  & 30.6 & 70.78 & 42.82 & 36.47 & 25.86 & 41.81 & 65.78 & 38.47 & 33.27 & 23.74 & 38.29 \\
		\cline{1-1}
		\cline{3-20}
		2 &  & 1x  & -  & - & - & - & - & \multirow{6}{*}{-} & \multirow{6}{*}{50.2} & 65.01 & 33.41 & 26.53 & 14.17 & 32.27 & 60.02 & 29.73 & 22.67 & 14.07 & 29.42 \\
		3 &  & - & 1x & - & - & - & - &  &  & 64.22 & 32.13$^\dag$ & 23.06$^\dag$ & 13.13$^\dag$ & 30.77$^\dag$ & 59.39 & 29.79 & 21.94$^\dag$ & 13.94 & 29.16$^\dag$ \\
		4  &  & -  & -  & 1x  & - & - & - &   &   & 63.30$^{\dag\star}$ & 32.61$^\dag$ & 25.16$^\dag$ & 13.65$^\dag$ & 31.25$^\dag$ & 58.18$^{\dag\star}$ & 30.92 & 21.33$^\dag$ & 14.12 & 29.08$^\dag$ \\
		5  &  & - & - & - & 1x & - & - &  &  & 62.89$^{\dag\star}$ & 32.06$^\dag$ & 22.67$^\dag$ & 13.88 & 30.62$^\dag$ & 57.20$^{\dag\star}$ & 28.83 & 20.98$^\dag$ & 14.15 & 28.51$^{\dag\star}$ \\
		6  &  & - & - & - & - & 1x & -  &   &  & 64.37 & 31.78$^\dag$ & 23.88$^\dag$ & 13.64$^\dag$ & 31.03$^\dag$ & 59.79 & 29.01 & 21.08$^{\dag\star}$ & 13.39$^{\dag\star}$ & 28.69$^{\dag\star}$ \\
		7  &  & - & - & - & - & - & 1x &   &   & 63.30$^{\dag\star}$ & 32.29$^\dag$ & 23.90$^\dag$ & 13.03$^\dag$ & 30.73$^\dag$ & 60.32 & 29.41 & 22.47$^\dag$ & 13.14$^{\dag\star}$ & 29.06$^\dag$ \\
		\cline{1-1}
		\cline{3-20}
		8 &  &  1x  & - & - & - & - & -  &  \multirow{6}{*}{2x} & \multirow{6}{*}{87.5}  & 62.64 & 30.25 & 23.84 & 13.68 & 30.28 & 61.14 & 28.47 & 21.24 & 12.92 & 28.74 \\
		9  &  & - & 1x & -  & -  & - & -  &  &   & 62.76 & 31.71 & 24.16 & 13.62 & 30.72 & 61.09 & 29.06 & 21.14 & 12.66 & 28.76 \\
		10  &  & -  & - & 1x & - & -  & - &   &   & 60.77$^{\dag\star}$ & 30.57$^{\star}$ & 23.45 & 13.86 & 29.91$^{\dag\star}$ & 59.57$^{\dag\star}$ & 29.06 & 20.20$^{\dag\star}$ & 13.36 & 28.44$^{\dag\star}$ \\
		11  &  & - & - & -  & 1x & -  & - &   &   & 60.86$^{\dag\star}$ & 30.81$^{\star}$ & 22.86$^{\dag\star}$ & 13.14$^{\dag\star}$ & 29.65$^{\dag\star}$ & 57.97$^{\dag\star}$ & 29.38 & 20.84 & 12.86 & 28.15$^{\dag\star}$ \\
		12  &  & - & - & -  & - & 1x  & - &   &   &  63.01 & 31.05$^{\star}$ & 23.10$^{\star}$ & 13.19$^{\dag\star}$ & 30.26$^{\star}$ & 60.18$^{\dag\star}$ & 28.74 & 20.41$^{\dag\star}$ & 12.91 & 28.43$^{\star}$ \\
		13  &  & - & - & -  & - & -  & 1x &   &   & 61.35$^{\dag\star}$ & 30.73$^{\star}$ & 22.14$^{\dag\star}$ & 12.78$^{\dag\star}$ & 29.49$^{\dag\star}$ & 60.50 & 28.92 & 20.43$^{\dag\star}$ & 12.34$^\dag$ & 28.33$^{\dag\star}$ \\
		\cline{1-1}
		\cline{3-20}
		14  &  & 2x  & -  & - & - & -  & - &  \multirow{6}{*}{2x} & \multirow{6}{*}{130.1}  & 62.69 & 31.49 & 24.47 & 13.90 & 30.79 & 61.09 & 29.62 & 21.31 & 13.67 & 29.28 \\
		15  &  & -  & 2x & - & -  & - & -  &   &   & 62.55 & 31.97 & 23.12$^\dag$ & 13.13 & 30.56 & 60.23 & 29.02 & 20.12$^\dag$ & 12.52$^\dag$ & 28.36$^\dag$ \\
		16  &  & -  & - & 2x & - & -  & - &  &  & 62.78 & 32.00 & 23.31$^\dag$ & 14.04 & 30.77 & 59.86$^{\dag\star}$ & 29.22 & 20.61 & 13.89 & 28.84$^\dag$ \\
		17  &  & -  & - & -  & 2x  & -  & - &   &  & 60.30$^{\dag\star}$ & 31.49 & 23.16$^\dag$ & 13.62 & 29.92$^{\dag\star}$ & 57.84$^{\dag\star}$ & 29.66 & 20.45$^\dag$ & 12.89$^\dag$ & 28.09$^{\dag\star}$ \\
		18  &  & -  & - & -  & -  & 2x & -  &   &   & 62.73 & 31.72 & 23.65 & 14.33 & 30.86 & 60.78 & 29.69 & 20.41$^\dag$ & 12.74$^\dag$ & 28.74$^\dag$ \\
		19  &  & -  & - & -  & - & -  & 2x &   &   & 61.98 & 31.37 & 23.06$^\dag$ & 13.39 & 30.18$^{\dag\star}$ & 59.18$^{\dag\star}$ & 28.79$^{\dag\star}$ & 19.71$^\dag$ & 12.26$^\dag$ & 27.85$^{\dag\star}$ \\
		\hline
		\hline
        20 & \multirow{7}{*}{\makecell[c]{Conformer\\+SpecAugment\\(19M)}} & \multicolumn{6}{c|}{-} & - & 30.6 & 82.90 & 63.60 & 57.47 &  47.28 & 61.01 & 78.69 & 58.80 & 55.62 & 45.61 & 57.96 \\
        \cline{1-1}
		\cline{3-20}
        21 &  & 2x & - & - & - & - & -  & \multirow{6}{*}{2x}  & \multirow{6}{*}{130.1} & 79.26 & 63.96 & 57.33 & 45.65 & 59.76 & 78.68 & 61.27 & 55.82 & 45.09 & 58.46 \\
		22 &  & - & 2x & - & - & - & -  &  &  & 66.77$^\dag$ & 49.39$^\dag$ & 46.47$^\dag$ & 42.02$^\dag$ & 50.03$^\dag$ & 66.50$^\dag$ & 48.72$^\dag$ & 46.84$^\dag$ & 41.72$^\dag$ & 49.76$^\dag$ \\
        23 &  & - & - & 2x & - & - & -  &  &  & 84.81 & 64.63 & 59.60 & 46.26 & 61.75 & 84.58 & 62.80 & 57.41 & 44.88 & 60.34 \\
		24 &  & -  & -  & - & 2x & - & -  &  &  & 66.47$^{\dag\star}$ & 48.17$^{\dag\star}$ & 46.43$^\dag$ & 42.26$^\dag$ & 49.70$^{\dag\star}$ & 65.93$^{\dag\star}$ & 49.33$^\dag$ & 46.78$^{\dag\star}$ & 41.66$^\dag$ & 49.77$^\dag$ \\
		25 &  & -  & -  & - & - & 2x & -  &  &  & 66.16$^{\dag\star}$ & 48.30$^{\dag\star}$ & 46.64$^\dag$ & 41.66$^\dag$ & 49.53$^{\dag\star}$ & 66.07$^\dag$ & 48.85$^\dag$ & 46.15$^{\dag\star}$ & 41.80$^\dag$ & 49.60$^\dag$ \\
		26 &  & -  & -  & - & - & - & 2x  &  &  & 65.34$^{\dag\star}$ & 47.87$^{\dag\star}$ & 46.54$^\dag$ & 41.98$^\dag$ & 49.33$^{\dag\star}$ & 65.13$^{\dag\star}$ & 48.17$^\dag$ & 46.05$^{\dag\star}$ & 41.85$^\dag$ & 49.23$^{\dag\star}$ \\
		\hline
		\hline
  	\end{tabular}
  	}
\end{table*}

\begin{table}[!t]
	\centering
	\caption{A comparison between published systems on UASpeech and our system. ``DA'' stands for data augmentation. ``L'', ``VL'' and ``Avg.'' represent WER (\%) for low, very low intelligibility group and average WER. }
	\label{tab:ua_comparison}
	\renewcommand\arraystretch{1.0}
    \renewcommand\tabcolsep{1.0pt}
	\scalebox{0.89}{
	\begin{tabular}{c|c|c|c}
	    \hline
	    \hline
		Systems & VL & L & Avg. \\
		\hline
		CUHK-2018 DNN System Combination \cite{yu2018development} & - & - & 30.60 \\
		Sheffield-2019 Kaldi TDNN + DA \cite{xiong2019phonetic} & 67.83 & 27.55 & 30.01 \\
		Sheffield-2020 Fine-tuning CNN-TDNN speaker adaptation \cite{xiong2020source} & 68.24 & 33.15 & 30.76 \\
		CUHK-2020 DNN + DA + LHUC-SAT \cite{geng2021investigation} & 62.44 & 27.55 & 26.37 \\
		CUHK-2021 LAS + CTC + Meta Learning + SAT \cite{wang2021improved} & 68.70 & 39.00 & 35.00 \\
		CUHK-2021 QuartzNet + CTC + Meta Learning + SAT \cite{wang2021improved} & 69.30 & 33.70 & 30.50 \\
		CUHK-2021 DNN + DCGAN + LHUC-SAT \cite{jin21_interspeech} & 61.42 & 27.37 & 25.89 \\
		CUHK-2021 DA + SBE Adapt + LHUC-SAT \cite{geng21b_interspeech} & 59.83 & 27.16 & 25.60 \\
		\textbf{TDNN + spectral basis GAN + LHUC-SAT (Sys. 19, Tab. \ref{tab:uaspeech})} & \textbf{59.18} & 28.79 & 27.85 \\
		\hline
		\hline
  	\end{tabular}
	}
\end{table}

In this experiment section, the performance of the proposed adversarial data augmentation approaches of Sec. \ref{sec:adv_parallel} and Sec. \ref{sec:adv_non_parallel} are investigated on four tasks: the English UASpeech \cite{kim2008dysarthric} and TORGO \cite{rudzicz2012torgo} dysarthric speech corpora as well as the English DementiaBank Pitt \cite{becker1994natural} and Cantonese JCCOCC MoCA \cite{xu2021speaker} elderly speech datasets. 
The baseline data augmentation method features both the standard speaker independent speed perturbation \cite{ko2015audio} of dysarthric or elderly speech and speaker dependent speed perturbation of control healthy or non-aged speech following our previous works \cite{geng2021investigation, ye2021development, liu2021recent}, for all four tasks. 
A range of acoustic models that give state-of-the-art performance on these tasks are chosen as the baseline speech recognition systems, including hybrid lattice-free maximum mutual information (LF-MMI) trained time delay neural network (TDNN) \cite{peddinti2015time, povey2016purely} and end-to-end (E2E) Conformer \cite{gulati2020conformer} models using additional online data augmentation via SpecAugment \cite{park19e_interspeech}. 
Model based speaker adaptation using learning hidden unit contribution (LHUC) \cite{swietojanski2016learning} is further applied. Sec. \ref{subsec:exp_dys} presents the experiments on the two dysarthric speech corpora while Sec. \ref{subsec:exp_eld} introduces experiments on the two elderly speech datasets. 
For all the speech recognition results measured in word error rate (WER) presented in this paper, matched pairs sentence-segment word error (MAPSSWE) based statistical significance test \cite{gillick1989some} is performed at a significance level $\alpha = 0.05$. 

\subsection{Experiments on Dysarthric Speech}
\label{subsec:exp_dys}

\subsubsection{the UASpeech Corpus}
The UASpeech corpus is the largest publicly available and widely used dysarthric speech dataset \cite{kim2008dysarthric}. It is an isolated word recognition task containing approximately $103$ hours of speech recorded from $29$ speakers, among whom $16$ are dysarthric speakers and $13$ are control healthy speakers. It is further divided into 3 blocks Block 1 (B1), Block 2 (B2) and Block 3 (B3) per speaker, each containing the same set of $155$ common words and a different set of $100$ uncommon words. The data from B1 and B3 of all the $29$ speakers are treated as the training set which contains $69.1$ hours of audio and $99195$ utterances in total, while the data from B2 collected of all the $16$ dysarthric speakers (excluding speech from control healthy speakers) are used as the test set containing $22.6$ hours of audio and $26520$ utterances in total. 

After removing excessive silence at both ends of the speech audio segments using an HTK \cite{young2002htk} trained GMM-HMM system, a combined total of $30.6$ hours of audio data from B1 and B3 ($99195$ utterances) are used as the training set, while $9$ hours of speech from B2 ($26520$ utterances) are used for performance evaluation. 
Data augmentation featuring speed perturbation of both the dysarthric speech in a speaker independent manner \cite{ko2015audio}, and the control healthy speech in a dysarthric speaker dependent fashion is further conducted \cite{geng2021investigation} to produce a $130.1$ hours augmented training set ($399110$ utterances, perturbing both healthy and dysarthric speech). 
If perturbing dysarthric data only, the resulting augmented training set contains $65.9$ hours of speech ($204765$ utterances).

\subsubsection{the TORGO Corpus}

The TORGO \cite{rudzicz2012torgo} corpus is a dysarthric speech dataset containing $8$ dysarthric and $7$ control healthy speakers with a total of approximately $13.5$ hours of audio data ($16394$ utterances). 
It consists of two parts: $5.8$ hours of short sentence based utterances and $7.7$ hours of single word based utterances. 
Similar to the setting of the UASpeech corpus, a speaker-level data partition is conducted by combining all $7$ control healthy speakers' data and two-thirds of the $8$ dysarthric speakers' data into the training set ($11.7$ hours). 
The remaining one-third of the dysarthric speech is used for evaluation ($1.8$ hours).  
After removal of excessive silence, the training and test set contain $6.5$ hours ($14541$ utterances) and $1$ hour ($1892$ utterances) of speech respectively. 
After data augmentation with both speaker dependent and speaker independent speed perturbation \cite{geng2021investigation,hu2022exploit}, the augmented training set contains $34.1$ hours of data ($61813$ utterances).

\subsubsection{Baseline ASR System Description}
For the UASpeech dataset, hybrid LF-MMI factored time delay neural network (TDNN) systems \cite{peddinti2015time, povey2016purely} containing $7$ context slicing layers are trained following the Kaldi \cite{povey2011kaldi} chain system setup, except that i-Vector features are not incorporated. 
The end-to-end (E2E) Conformer systems are implemented using the ESPnet toolkit \cite{watanabe2018espnet} \footnote{$8$ encoder layers + $4$ decoder layers, feed-forward layer dim = $1024$, attention heads = $4$, dim of attention heads = $256$, interpolated CTC+AED cost.} to directly model grapheme (letter) sequence outputs. 
$80$-dimensional Mel-scale filter-bank input features are used in the E2E Conformer systems, while $40$-dimensional Mel-scale filter-bank input features and a $3$-frame context window is used in the hybrid TDNN system.
Following the configurations given in \cite{christensen2012comparative,yu2018development}, a uniform language model with a word grammar network is used in decoding.

On the TORGO dataset, the hybrid LF-MMI TDNN and E2E graphemic Conformer systems use the same configurations as those adopted above for the UASpeech data, except that $40$-dimensional Mel-scale filter-bank input features are used for both systems. 
A $3$-gram language model (LM) trained by all the TORGO transcripts with a vocabulary size of $1.6$k is used during recognition with both the hybrid TDNN and E2E Conformer systems. 

\subsubsection{Performance of Data Augmentation on Dysarthric Speech}

\begin{table}[!t]
	\centering
	\caption{Performance of GAN data augmentation approaches on expanded training data before and after LHUC-SAT speaker adaptation on the \textbf{TORGO} \cite{rudzicz2012torgo} test set. ``SE./MOD./Mild'' refers to the speech impairment severity levels: severe, moderate and mild. $\dag$ denotes a statistically significant improvement ($\alpha=0.05$) is obtained over the comparable baseline systems (Sys. 2, 6, 10). Other naming conventions are the same as those in Tab. \ref{tab:uaspeech} for UASpeech.}
	\label{tab:torgo}
	\renewcommand\arraystretch{1.0}
    \renewcommand\tabcolsep{1.4pt}
	\scalebox{0.88}{
	\begin{tabular}{c|c|c|c|c|c|c|c|c|c}
	    \hline
	    \hline
		\multirow{3}{*}{Sys.} & \multirow{3}{*}{\makecell[c]{Model \\(\# Parameters)}} & \multicolumn{2}{c|}{Data Aug.} & \multirow{3}{*}{\# Hrs.} & \multirow{3}{*}{\makecell[c]{LHUC\\SAT}} & \multicolumn{4}{c}{WER \% }  \\
		\cline{3-4}
		\cline{7-10}
		& & \multirow{2}{*}{CTRL} & DYS &  &  & \multirow{2}{*}{SE.} & \multirow{2}{*}{MOD.} & \multirow{2}{*}{Mild} & \multirow{2}{*}{Avg.} \\
		\cline{4-4}
		&  &  & S &  &  &  &  &  &  \\
		\hline
		\hline
		1 & \multirow{9}{*}{\makecell[c]{Hybrid\\TDNN\\(10M)}} & \multicolumn{2}{c|}{-} & 6.5 & - & 16.22 & 10.31 & 3.87 & 11.62 \\
		\cline{1-1}
		\cline{3-10}
		2 &  & S & \multirow{8}{*}{\checkmark} & \multirow{8}{*}{34.1} & \multirow{4}{*}{-} & 12.80 & 8.78 & 3.64 & 9.47  \\
		3 &  & SG  &  &  &  & 13.90 & 5.31$^\dag$ & 3.02$^\dag$ & 9.15 \\
		4 &  & SBG &  &  &  & 13.29 & 5.82$^\dag$ & 2.86$^\dag$ & 8.90$^\dag$ \\
		5 &  & SBG+SG &  &  &  & 12.93 & 5.10$^\dag$ & 2.86$^\dag$ & 8.56$^\dag$  \\
		\cline{1-1}
		\cline{3-3}
		\cline{6-10}
		6 &  & S &  &  & \multirow{4}{*}{\checkmark}  & 12.52 & 8.27 & 3.25 & 9.11 \\
		7 &  & SG  &  &  &  & 13.90 & 5.00$^\dag$ & 3.10 & 9.11 \\
		8 &  & SBG &  &  &  & 13.66 & 4.90$^\dag$ & 2.55$^\dag$ & 8.81 \\
		9 &  & SBG+SG &  &  &  & 12.97 & 5.00$^\dag$ & 2.20$^\dag$ & 8.51$^\dag$ \\
		\hline
		\hline
		10 & \multirow{4}{*}{\makecell[c]{Conformer\\+SpecAugment\\(18M)}} & S & \multirow{4}{*}{\checkmark} & \multirow{4}{*}{34.1} & \multirow{4}{*}{-} & 21.66 & 6.22 & 4.10 & 13.67 \\
		11 &  & SG &  &  &  & 20.04 & 6.63 & 3.86 & 12.80 \\
		12 &  & SBG &  &  &  & 20.00$^\dag$ & 6.32 & 3.40 & 12.63$^\dag$ \\
		13 &  & SBG+SG &  &  &  & 20.44 & 5.81 & 4.17 & 12.97 \\
		\hline
		\hline
  	\end{tabular}
	}
\end{table}

The performance of various speaker dependent GAN based data augmentation approaches proposed in Sec. \ref{sec:adv_parallel} and Sec. \ref{sec:adv_non_parallel} evaluated on the UASpeech test set of 16 dysarthric speakers are shown in Tab. \ref{tab:uaspeech}.
We can observe the following trends from Tab. \ref{tab:uaspeech} on the hybrid TDNN systems: 

\noindent
\textbf{i)} Among the two adversarial data augmentation approaches introduced in Sec. \ref{sec:adv_parallel} requiring parallel control-dysarthric audio recordings, the tempo-GAN approach consistently outperforms the comparable baseline tempo perturbation (Sys. 4 \textit{vs.} 2, Sys. 10 \textit{vs.} 8 and Sys. 16 \textit{vs.} 14, col. 15 ``Avg.'' for unadapted systems) by up to 1.02\% absolute (3.16\% relative) average WER reduction (Sys. 4 \textit{vs.} 2, col. 15); 

\noindent
\textbf{ii)} The speed-GAN approach consistently outperforms the speed perturbation baselines (Sys. 5 \textit{vs.} 3, Sys. 11 \textit{vs.} 9 and Sys. 17 \textit{vs.} 15, col. 15) by up to 1.07\% absolute (3.56\% relative) average WER reduction (Sys. 11 \textit{vs.} 9, col. 15); 

\noindent
\textbf{iii)} Our proposed speed-GAN approach consistently outperforms the tempo-GAN approach (Sys. 5 \textit{vs.} 4, Sys. 11 \textit{vs.} 10 and Sys. 17 \textit{vs.} 16) by up to 0.85\% absolute (2.76\% relative) average WER reduction (Sys. 17 \textit{vs.} 16, col. 15);

\noindent
\textbf{iv)} The spectral basis GAN based augmentation method introduced in Sec. \ref{sec:adv_non_parallel} can be applied to both parallel and non-parallel data, by performing the adversarial neural transformation to the spectral basis vectors of any pair of control and target speaker's dysarthric utterances with or without the same spoken contents. 
In the UASpeech experiments where parallel control-dysarthric speech recordings of identical word contents are provided, the source and target spectral basis vectors are extracted from such parallel control-dysarthric utterances. 
The resulting ``SBG'' augmented data trained TDNN systems (Sys. 6, 12, 18) produced performance comparable to the baseline temporal (Sys. 2, 8, 14) and speed perturbation (Sys. 3, 9, 15) methods. 
When the speaker dependent spectral basis GAN (SBG) generated data is further speed perturbed and transformed using the speed-GAN (SG) model in a pipelined two-stage manner, leading to the ``SBG+SG'' systems, further WER reductions are obtained (Sys. 7, 13, 19 \textit{vs.} Sys. 6, 12, 18). 
Among these systems, Sys. 19 also gives the lowest WER on the ``Very Low'' intelligibility group of UASpeech as summarized in Tab. \ref{tab:ua_comparison} when contrasted with a set of recently published results on UASpeech.

\noindent
\textbf{v)} Similar trends are retained after LHUC-SAT speaker adaptation is applied (last column for LHUC-SAT adapted systems). 
On the largest $130.1$ hour augmented training set, the pipelined ``SBG+SG'' augmentation system (Sys. 19) produces the lowest average WER among all systems using the same amount of augmented training data (Sys. 14 to 19, last column).

Similar trends are also observed on the end-to-end Conformer systems (with SpecAugment\cite{park19e_interspeech} applied) where the proposed speed-GAN (SG), spectral basis GAN (SBG) augmentation approaches and their combination (SBG+SG) consistently produce statistically significant WER reductions over the baseline tempo or speech perturbation (Sys. 24, 25, 26 \textit{vs.} Sys. 21 \& 22) over the ``Very Low'' (VL) and ``Low'' (L) intelligibility subsets. 

A similar set of experiments are then conducted using the TORGO data and shown in Tab. \ref{tab:torgo} where the contrast is drawn between the baseline speed perturbation and the adversarial data augmentation approaches: speed-GAN (SG), spectral basis GAN (SBG) and their pipelined combination (SBG+SG). 
Trends similar to those found on the UASpeech data of Tab. \ref{tab:uaspeech} can be found.

\noindent
\textbf{i)} All the three speaker dependent GAN based data augmentation approaches consistently outperform the comparable baseline system using both speaker independent and dependent speed perturbation (Sys. 3, 4, 5 \textit{vs.} 2) by up to {\bf 0.91\% absolute (9.61\% relative)} average WER reduction (Sys. 5 \textit{vs.} 2, last col.) before LHUC-SAT speaker adaptation; 

\noindent
\textbf{ii)} When further combined with LHUC-SAT, the two systems trained using spectral basis GAN based augmentation consistently outperform the baseline with speed perturbation only (Sys. 8, 9 \textit{vs.} 6) by up to 0.6\% absolute (6.59\% relative) average WER reduction (Sys. 9 \textit{vs.} 6, last col.); 

\noindent
\textbf{iii)} On the Conformer systems, GAN based data augmentation approaches also consistently outperform the baseline speed perturbation (Sys. 11, 12, 13 \textit{vs.} 10). 
Among the three GAN augmented Conformer systems, the spectral basis GAN based approach (Sys. 12) produces the lowest WER of 12.63\%, with a statistically significant WER reduction of 1.04\% over the baseline Conformer (Sys. 10) using speed perturbation only.

\begin{table*}[!t]
	\centering
	\caption{Performance of spectral basis GAN based data augmentation approaches on the expanded training data before and after LHUC-SAT speaker adaptation on the \textbf{DementiaBank Pitt} corpus \cite{becker1994natural} development (Dev.) and evaluating (Eval) set. 
	``INV'' and ``PAR'' in the ``Data Augmentation'' column refer to non-aged clinical investigator and elderly participant respectively. In the ``INV'' column, ``S'' denotes speed perturbation and ``SBG'' stands for spectral basis GAN. ``rand'', ``avg'' and ``exhaustive'' stand for the three spectral basis vectors pairing schemes between investigator and elderly speech in GAN training described in Sec. \ref{subsec:non_parallel_gan_arch}. Speaker independent speed perturbation is applied to elderly speech. $\dag$ denotes a statistically significant improvement ($\alpha= 0.05$) obtained over the comparable baseline systems (Sys. 3, 7, 11)}
	\label{tab:dbank}
 	\renewcommand\arraystretch{1.0}
    \scalebox{0.85}{
	\begin{tabular}{c|c|c|c|c|c|c|c|c|c|c}
	    \hline
	    \hline
		\multirow{3}{*}{Sys.} & \multirow{3}{*}{\makecell[c]{Model\\(\# Parameters)}} & \multicolumn{2}{c|}{Data Augmentation} & \multirow{3}{*}{\# Hrs.} & \multirow{3}{*}{\makecell[c]{LHUC\\SAT}} & \multicolumn{5}{c}{WER \% }  \\
		\cline{3-4}
		\cline{7-11}
		& & \multirow{2}{*}{INV} & \multirow{2}{*}{PAR} &  &  & \multicolumn{2}{c|}{Dev.}  &  \multicolumn{2}{c|}{Eval} & \multirow{2}{*}{Avg.} \\
		\cline{7-10}
		&  &  &  &  &  & INV & PAR & INV & PAR & \\
		\hline
		\hline
		1 & \multirow{10}{*}{\makecell[c]{Hybrid\\TDNN\\(18M)}} & \multirow{2}{*}{-} & \multirow{2}{*}{-} & \multirow{2}{*}{15.8} & - &  21.45 & 51.38 & 21.09 & 39.49 & 36.31 \\
		\cline{6-11}
		2 &  &  &  &  &  \checkmark & 21.03 & 50.29 & 21.75 & 38.46 & 35.55  \\
		\cline{1-1}
		\cline{3-11}
		3 &  & S & \multirow{8}{*}{\checkmark} & \multirow{8}{*}{58.9} & \multirow{4}{*}{-} & 19.91 & 47.93 & 19.76 & 36.66 & 33.80 \\
		4 &  & SBG (rand) &  &  &  & 19.27$_{\pm0.16}$ & 45.28$_{\pm0.52}$ & 18.95$_{\pm0.55}$ & 34.28$_{\pm0.30}$ & 32.16$_{\pm0.20}^\dag$  \\ 
		5 &  & SBG (avg) &  &  &  & 19.25 & 44.93 & 19.09 & 34.31 & 31.93$^\dag$ \\
		6 &  & SBG (exhaustive) &  &  &  & 19.41 & 46.13 & 19.42 & 35.02 & 32.60$^\dag$ \\
		\cline{1-1}
		\cline{3-3}
		\cline{6-11}
		7 &  & S &  &  & \multirow{4}{*}{\checkmark} & 19.26 & 45.49 & 18.42 & 35.44	 & 32.33 \\
		8 &  & SBG (rand) &  &  &  & 18.66$_{\pm0.01}$ & 45.14$_{\pm0.00}$ & 19.31$_{\pm0.04}$ & 34.12$_{\pm0.01}$ & 31.75$_{\pm0.00}^\dag$  \\
		9 &  & SBG (avg) &  &  &  & 18.62 & 44.88 & 18.65 & 34.08 & 31.60$^\dag$ \\ 
		10 &  & SBG (exhaustive) &  &  &  & 18.68 & 45.33 & 19.20 & 34.22 & 31.85$^\dag$ \\
		\hline
		\hline
		11 & \multirow{4}{*}{\makecell[c]{Conformer\\+SpecAugment\\(52M)}} & S & \multirow{4}{*}{\checkmark} & \multirow{4}{*}{58.9} & \multirow{4}{*}{-} & 20.80 & 47.85 & 19.86 & 36.09 & 34.00 \\
		12 &  & SBG (rand) &  &  &  & 22.23$_{\pm 0.06}$ & 45.22$_{\pm 0.02}^\dag$ & 20.05$_{\pm 0.48}$ & 35.06$_{\pm 0.04}$ & 33.42$_{\pm 0.03}^\dag$ \\
		13 &  & SBG (avg) &  &  &  & 21.85 & 46.29$^\dag$ & 19.42 & 34.97 & 33.65 \\
		14 &  & SBG (exhaustive) &  &  &  & 22.83 & 47.64 & 19.87 & 36.22 & 34.80 \\
		\hline
		\hline
  	\end{tabular}
    }
\end{table*}

\begin{table}[!t]
    \centering
    \caption{Ablation study on the effect of using different settings of the spectral basis GAN output perturbation scaling parameter $\lambda$ of Sec. \ref{subsec:non_parallel_gan_training} on the \textbf{DementiaBank Pitt} corpus \cite{becker1994natural} with $58.9$h augmented training data. Naming conventions are the same as those in Tab. \ref{tab:dbank}.}
    \label{tab:lambda_dbank}
    \scalebox{0.9}{
    \begin{tabular}{c|c|c|c|c|c}
    \hline
    \hline
    \multirow{2}{*}{$\lambda$} & \multicolumn{2}{c|}{Dev.} & \multicolumn{2}{c|}{Eval} & \multirow{2}{*}{Avg.} \\
    \cline{2-5}
    & INV. & PAR. & INV. & PAR. & \\
    \hline
    \hline
    0.001 & 19.75 & 46.84 & 19.64 & 34.85 & 33.00 \\
    0.01 & 19.35 & 45.78 & 18.20 & 34.24 &  32.27 \\
    0.1 & 19.52 & 46.58 & 19.87 & 34.71 & 32.79 \\
    0.2 & 19.25 & 44.93 & 19.09 & 34.31 & \textbf{31.93} \\
    1 & 19.10 & 45.70 & 18.76 & 33.49 & 32.03 \\
    2 & 19.23 & 45.67 & 18.87 & 34.20 & 32.19 \\
    5 & 19.71 & 45.99 & 17.87 & 34.39 & 32.51 \\
	\hline
	\hline
    \end{tabular}
    }
\end{table}

\subsection{Experiments on Elderly Speech}
\label{subsec:exp_eld}

\subsubsection{the DementiaBank Pitt Corpus}
The DementiaBank Pitt \cite{becker1994natural} corpus contains approximately $33$ hours of audio data recorded over interviews between the $292$ elderly participants and the clinical investigators. 
It is further split into a $27.2$h training set, a $4.8$h development and a $1.1$h evaluation set for ASR system development. 
The evaluation set is based on exactly the same $48$ speakers' Cookie (picture description) task recordings as those in the ADReSS \cite{luz2020alzheimer} test set, while the development set contains the remaining recordings of these speakers in other tasks if available. 
The training set contains $688$ speakers ($244$ elderly participants and $444$ investigators), while the development set includes $119$ speakers ($43$ elderly participants and $76$ investigators) and the evaluation set contains $95$ speakers ($48$ elderly participants and $47$ investigators).
Different sets of speakers are used in the training, development and evaluation sets. 
After removal of excessive silence \cite{ye2021development}, the training set contains $15.7$ hours of audio data ($29682$ utterances) while the development and evaluation sets contain $2.5$ hours ($5103$ utterances) and $0.6$ hours ($928$ utterances) of audio data respectively. 
Data augmentation featuring speaker independent speed perturbation of elderly speech and elderly speaker dependent speed perturbation of non-aged investigators' speech \cite{ye2021development} produced a $58.9$h augmented training set ($112830$ utterances).

\subsubsection{the JCCOCC MoCA Corpus}
The Cantonese JCCOCC MoCA corpus contains conversations recorded from cognitive impairment assessment interviews between $256$ elderly participants and the clinical investigators \cite{xu2021speaker}. 
The training set contains $369$ speakers ($158$ elderly participants and $211$ investigators) with a duration of $32.4$ hours. 
The development and evaluation sets each contains speech recorded from two different sets of $49$ elderly speakers that are not covered by the training set. 
After removal of excessive silence, the training set contains $32.1$ hours of speech ($95448$ utterances) while the development and evaluation sets contain $3.5$ hours ($13675$ utterances) and $3.4$ hours ($13414$ utterances) of speech respectively. 
After applying the same baseline speaker independent and dependent speed perturbation based data augmentation adopted above for the DementiaBank Pitt corpus, the expanded training set consists of $156.9$ hours of speech ($389409$ utterances). 
Different sets of speakers are used in the training, development and evaluation sets.

\subsubsection{Baseline ASR System Description} 
The hybrid LF-MMI TDNN and E2E graphemic Conformer systems use the same configurations as those adopted above for the UASpeech data, except that $40$-dimensional Mel-scale filter-bank input features are used.
On the English DementiaBank data, for both the hybrid TDNN and E2E graphemic Conformer systems, a word level $4$-gram LM with Kneser-Ney smoothing is trained using the SRILM toolkit \cite{stolcke2002srilm} following the settings of our previous work \cite{yu2018development} and a $3.8$k word recognition vocabulary covering all the words in the DementiaBank Pitt corpus is used in recognition. 
On the Cantonese JCCOCC MoCA data, the Conformer model training used Cantonese Characters as the output targets. 
A word level $4$-gram language model with Kneser-Ney smoothing is trained on the acoustic transcription ($610$k words) and a $5.2$k recognition vocabulary covering all the words in the JCCOCC MoCA corpus is also used.

\subsubsection{Performance of Data Augmentation on Elderly Speech}
The performance of various spectral basis GAN (SBG) based augmentation approaches proposed in Sec. \ref{sec:adv_non_parallel} for non-parallel data are evaluated on the DementiaBank Pitt test set is shown in Tab. \ref{tab:dbank}. 
First, the following trends can be found from Tab. \ref{tab:dbank} on the hybrid TDNN systems:

\noindent
\textbf{i)} Before LHUC-SAT speaker adaptation is applied, all the three SBD based data augmentation approaches with three different spectral bases pairing schemes between non-aged investigator and elderly participant speech described in Sec. \ref{subsec:parallel_gan_arch} consistently outperform the comparable baseline TDNN system using both speaker independent and dependent speed perturbation (Sys. 4, 5, 6 \textit{vs.} 3). 
Among these three pairing schemes, using the average, mean spectral bases computed over all the target elderly speaker's utterances (Sys. 5, ``SBG (avg)'') as the GAN training targets gives the largest improvement by up to 1.87\% absolute (5.53\% relative) average WER reduction (Sys. 5 \textit{vs.} 3, last col.). 
More specifically, on the two elderly speech ``PAR'' subsets statistically significant WER reductions of {\bf 2.35\%-3.0\% absolute (6.3\%-6.4\% relative)} are obtained over the baseline speed perturbation (Sys. 5 \textit{vs.} 3, col. 8 \& 10). 
For the SGB systems using a random pairing (Sys. 4, 8, ``SBG (rand)''), a total of 10 TDNN systems are trained using 10 different investigator-participant utterance spectral bases paring before computing both the WER mean and standard deviation (shown after ``$\pm$''). 

\noindent
\textbf{ii)} When further combined with LHUC-SAT speaker adaptation, the three systems trained with spectral basis GAN based augmentation again consistently outperform the baseline with speed perturbation only (Sys. 8, 9, 10 \textit{vs.} 7) by up to 0.73\% absolute (2.26\% relative) average WER reduction (Sys. 9 \textit{vs.} 7, last col.). 
The reduced performance gap between the baseline system using speed perturbation and those using spectral basis GANs after LHUC-SAT adaptation suggest the GAN augmented data produced better coverage of speaker level heterogeneity in the speaker independent TDNN system training before adaptation to unseen speakers in the test data, but not covered in the DementiaBank Pitt training data set.

\noindent
\textbf{iii)} A further ablation study on the effect of using different settings of the spectral basis GAN output perturbation scaling parameter $\lambda$ of Sec. \ref{subsec:parallel_gan_training} is then conducted using the ``SBG (avg)'' augmented TDNN systems and their performance are shown in Tab. \ref{tab:lambda_dbank}.
An optimal setting of $\lambda = 0.2$ is used in all the experiments of Tab. \ref{tab:dbank} and the following Cantonese elderly speech experiments of Tab. \ref{tab:jcmoca}.

Second, on the Conformer systems, the proposed spectral basis GAN trained using the randomly paired spectral bases ``SBG (rand)'' and mean spectral bases ``SBG (mean)'' outperforms the baseline speed perturbation method (Sys. 12, 13 \textit{vs.} 11) by up to 0.58\% absolute (1.7\% relative) CER deduction (Sys. 12 \textit{vs.} 11).
Among the two GAN augmented Conformer systems (Sys. 12 \& 13), the spectral basis GAN trained using the randomly paired spectral bases of target elderly speaker (Sys. 12, ``SBG (rand)'') as the GAN training targets produces the lowest average CER of 33.42\%. 

\begin{table}[!t]
	\centering
	\caption{Performance of spectral basis GAN based data augmentation approaches on the expanded training data before and after LHUC-SAT speaker adaptation on the \textbf{JCCOCC MoCA} corpus \cite{xu2021speaker} development (Dev.) and evaluation (Eval) sets containing elderly speakers only. $\dag$ denotes a statistically significant improvement ($\alpha = 0.05$) obtained over the comparable baseline systems (Sys. 3, 6, 9). Other naming conventions are the same as those in Tab. \ref{tab:dbank} for the DementiaBank Pitt data.}
	\label{tab:jcmoca}
	\renewcommand\arraystretch{1.0}
    \renewcommand\tabcolsep{1.1pt}
	\scalebox{0.82}{
	\begin{tabular}{c|c|c|c|c|c|c|c|c}
	    \hline
	    \hline
		\multirow{2}{*}{Sys.} & \multirow{2}{*}{\makecell[c]{Model\\(\# Parameters)}} & \multicolumn{2}{c|}{Data Aug.} & \multirow{2}{*}{\# Hrs.} & \multirow{2}{*}{\makecell[c]{LHUC\\SAT}} & \multicolumn{3}{c}{CER \%} \\
		\cline{3-4}
		\cline{7-9}
		 & & INV & PAR &  &  & Dev.  &  Eval & Avg. \\
		\hline
		\hline
		1 & \multirow{8}{*}{\makecell[c]{Hybrid\\TDNN\\(18M)}} & \multirow{2}{*}{-} & \multirow{2}{*}{-} & \multirow{2}{*}{32.1} & - & 30.89 & 27.85 & 29.36  \\
		\cline{6-9}
		2 &  &  &  &  & \checkmark & 30.05 & 27.34 & 28.69  \\
		\cline{1-1}
		\cline{3-9}
		3 &  & S & \multirow{3}{*}{\checkmark} & \multirow{3}{*}{156.9} & \multirow{3}{*}{-} & 26.87 & 23.71 & 25.28  \\
		4 &  & SBG (rand) &  &  &  & 25.96$_{\pm0.02}^\dag$ & 22.86$_{\pm0.02}^\dag$ & 24.40$_{\pm0.02}^\dag$   \\
		5 &  & SBG (avg) &  &  &  & 25.88$^\dag$ & 22.86$^\dag$ & 24.36$^\dag$   \\
		\cline{1-1}
		\cline{3-9}
		6 &  & S & \multirow{3}{*}{\checkmark} & \multirow{3}{*}{156.9} & \multirow{3}{*}{\checkmark} & 25.77 & 22.94 & 24.35  \\
		7 &  & SBG (rand) &  &  &  & 24.82$_{\pm0.20}^\dag$ & 22.08$_{\pm0.11}^\dag$ & 23.44$_{\pm0.12}^\dag$   \\
		8 &  & SBG (avg) &  &  &  & 24.69$^\dag$ & 22.12$^\dag$ &  23.40$^\dag$ \\ 
		\hline
		\hline
		9 & \multirow{3}{*}{\makecell[c]{Conformer\\+SpecAugment\\(53M)}} & S & \multirow{3}{*}{\checkmark} & \multirow{3}{*}{156.9}  & \multirow{3}{*}{-} & 33.08 & 31.24 & 32.15  \\
		10 &  & SBG (rand) &  &  &  & 32.18$^\dag_{\pm0.05}$ & 30.28$^\dag_{\pm0.07}$ & 31.22$^\dag_{\pm0.05}$  \\
		11 &  & SBG (avg) &  &  &  & 31.78$^\dag$ & 30.35$^\dag$ & 31.06$^\dag$  \\
		\hline
		\hline
  	\end{tabular}
	}
\end{table}

A similar set of experiments are then conducted using the Cantonese JCCOCC MoCA data and are shown in Tab. \ref{tab:jcmoca}.  
Several trends similar to those observed on the DementiaBank Pitt data of Tab. \ref{tab:dbank} can be found. 

\noindent
\textbf{i)} The SBG based data augmentation approaches using either the mean of, 
or randomly selected, 
target elderly speaker's utterance level spectral bases in training consistently outperform the baseline TDNN system using speed perturbation (Sys. 4, 5 \textit{vs.} 3) by up to 0.92\% absolute (3.64\% relative) average character error rate (CER) reduction (Sys. 5 \textit{vs.} 3, last col.); 

\noindent
\textbf{ii)} When further applying LHUC-SAT speaker adaptation, the same two systems trained with SBG augmentation again outperform the baseline system with speed perturbation (Sys. 7, 8 \textit{vs.} 6) by up to 0.95\% absolute (3.90\% relative) average CER reduction (Sys. 8 \textit{vs.} 6, last col.); 

\noindent
\textbf{iii)} The proposed spectral basis GAN augmentation also outperform the speed perturbation method on the Conformer systems (Sys. 10, 11 \textit{vs.} 9) by up to 1.0\% absolute (3.39\% relative) statistically significant CER reduction (Sys. 11 \textit{vs.} 9, last col.). 
Among the two GAN augmented Conformer systems, the spectral basis GAN trained using the mean spectral bases of each elderly speaker (Sys. 11, ``SBG (avg)'') produces the lowest CER of 31.06\%. 

\section{Conclusion}
\label{sec:conclusion}
This paper presents a set of speaker dependent generative adversarial networks (GAN) based data augmentation approaches for dysarthric and elderly speech recognition tasks. 
These personalized adversarial data augmentation approaches can flexibly expand the limited amounts of dysarthric or elderly training data that traditionally hinders ASR system development for these atypical speech domains with or without the use of parallel audio recordings. 
Experimental results obtained on state-of-the-art hybrid TDNN and end-to-end Conformer ASR systems evaluated across four dysarthric or elderly speech datasets of two languages suggest improved coverage in the resulting augmented data and model generalisation are obtained over the baseline systems using conventional methods based on tempo or speed perturbation and SpecAugment. 
Future research will focus on improving adversarial data augmentation techniques to inject richer spectral and temporal characteristics of dysarthric and elderly speech.


\bibliography{mybib}
\bibliographystyle{IEEEtran.bst}

\newpage

 





\end{document}